\documentclass[useAMS,usenatbib]{mnras}
\usepackage{amsmath}
\usepackage{amssymb}
\usepackage{graphicx}
\graphicspath{{./figures/}}
\usepackage{hyperref}
\usepackage{pdfpages}

\newcommand{\eq}[1]{\begin{equation}\begin{split}#1\end{split}\end{equation}}

\newcommand{\eal}[1]{\begin{align}#1\end{align}}

\begin{document}

\title[Analytical Theory of Second-Order Resonances]{Migration of Planets Into and Out of Mean Motion Resonances in 
Protoplanetary Disks: Analytical Theory of Second-Order Resonances}

\author[Xu and Lai]{Wenrui Xu,$^{1}$ Dong Lai$^{1,2}$\\
$^{1}$Cornell Center for Astrophysics and Planetary Science, Department of Astronomy, Cornell University, Ithaca, NY 14853, USA\\
$^{2}$Institute for Advanced Study, Princeton, NJ 08540, USA}

\maketitle

\begin{abstract}
Recent observations of {\it Kepler} multi-planet systems have revealed
a number of systems with planets very close to second-order mean
motion resonances (MMRs, with period ratio $1:3$, $3:5$, etc.) We
present an analytic study of resonance capture and its stability for
planets migrating in gaseous disks. Resonance capture requires slow
convergent migration of the planets, with sufficiently large
eccentricity damping timescale $T_e$ and small pre-resonance
eccentricities. We quantify these requirements and find that they can
be satisfied for super-Earths under protoplanetary disk
conditions. For planets captured into resonance, an equilibrium state
can be reached, in which eccentricity excitation due to resonant
planet-planet interaction balances eccentricity damping due to
planet-disk interaction. This ``captured'' equilibrium
can be overstable, leading to partial or permanent escape of the
planets from the resonance. In general, the stability of the captured
state depends on the inner to outer planet mass ratio $q=m_1/m_2$ and
the ratio of the eccentricity damping times. The overstability growth
time is of order $T_e$, but can be much larger for systems close to
the stability threshold. For low-mass planets undergoing type I
(non-gap opening) migration, convergent migration requires $q \lesssim 1$, 
while the stability of the capture requires $q\gtrsim 1$. These
results suggest that planet pairs stably captured into second-order
MMRs have comparable masses. This is in contrast to first-order MMRs,
where a larger parameter space exists for stable resonance capture. We
confirm and extend our analytical results with $N$-body simulations,
and show that for overstable capture, the escape time from the MMR can
be comparable to the time the planets spend migrating between
resonances.
\end{abstract}


\section{Introduction}

Planets formed in a gaseous protoplanetary disk excite density waves
and experience back-reaction torques from the disk 
\citep{GoldreichTremaine79,GoldreichTremaine80,LinPapaloizou79}.
While the magnitude and sign of the torque 
depend on the planet's mass, location and the physical property of the disk 
(e.g., \citealt{KleyNelson12}; \citealt{Baruteau14}),
some degrees of disk-driven migration are inevitable, especially for planets with
gaseous atmosphere or envelope (including gas giants and ``rocky'' planets
with radii $\gtrsim 2R_\oplus$) -- the presence of the envelope 
indicates that these planets have formed before gas disks dissipate. 
Planets in multi-planet system generally have different migration
rates, and their period ratio varies during migration. It has
long been established that slow convergent migration can naturally result in
neighboring planets captured into mean motion resonances (MMRs), in
which the planet's period ratio $P_2/P_1$ stays close to $j/(j-q)$ (where $q,j>q$
are positive integers) (e.g. \citealt{Snellgrove01}; \citealt{LeePeale02};
see also \citet{Goldreich65} and \citet{Peale86} for studies of MMRs in the solar system).
Therefore one would expect that an appreciable fraction of multi-planet systems
reside in resonant configurations.

The {\it Kepler} mission has discovered thousands of super-Earths and
sub-Neptunes (with radii 1.2-3$R_\oplus$) with periods less than
200~days, many of which are in multi-planet systems.  The period
ratios of a majority of these {\it Kepler} multi's do not
preferentially lie in or close to MMRs, although there is a
significant excess of planet pairs with period ratios slightly larger
(by about $1-2\%$) than that for exact resonance \citep{Fabrycky14}.
The discovery of several resonant chain systems (such as 
Kepler-223, with four planets in 3:4:6:8 MMRs; \citealt{Mills16}) 
suggests that resonance capture during the ``clean'' disk-driven migration phase
can be quite common, but subsequent physical processes may have destroyed
the resonances for most systems. There have been many numerical studies
on MMRs in protoplanetary disks, either 
using $N$-body integrations with fictitious forces that mimic dissipative effects
(e.g., \citealt{LeePeale02}; \citealt{TerquemPapaloizou07}; \citealt{ReinPapaloizou09};
\citealt{Rein12}; \citealt{Migaszewski15})
or using self-consistent hydrodynamics (e.g.
\citealt{Kley05}; \citealt{PapaloizouSzuszkiewicz05}; \citealt{Crida08}; \citealt{Zhang14}; \citealt{AndrePapaloizou16}).
A number of papers have studied possible mechanisms
that could break migration-induced resonances, such as
late dynamical instabilities following disk dispersal 
(e.g., \citealt{Cossou14}; \citealt{PuWu15}),
planetesimal scattering \citep{ChatterjeeFord15},
and tidal dissipation in the planets
\citep{PapaloizouTerquem10,LithwickWu12,BatyginMorbidelli13,Lee13}.
Regardless of whether MMRs are maintained or destroyed by any of these processes, 
it is important to recognize that MMRs play a significant role in 
the early evolution of planetary systems and can profoundly shape their final 
architectures.  Therefore it is important to understand under what conditions
resonance capture can occur or can be avoided when planets undergo 
disk-driven migration.

Recent theoretical works on resonance capture have focused on
first-order ($j:j-1$) MMRs. While the basic dynamics of resonance capture
was long understood (i.e., convergent migration with sufficiently slow
rate leads to either secure or probabilistic resonance capture,
depending on the initial, ``pre-resonance'' planet eccentricity;
see, e.g., \citealt{Henrard82,BorderiesGoldreich84,Lemaitre1984a,Lemaitre1984b}),
a recent study by \citet{GoldreichSchlichting14} revealed that, in the
presence of the eccentricity damping due to planet-disk interaction, the
long-term stability of MMRs can be compromised and resonance capture
may be temporary.  In particular, using a restricted three-body model
(where the inner planet has a negligible mass), \citet{GoldreichSchlichting14}
 showed that the equilibrium MMR state of the planets
(in which eccentricity excitation due to resonant
planet-planet interaction balances disk-induced eccentricity damping) can be overstable,
and the system escapes the resonance on
a timescale shorter than the migration timescale between resonances. 
A more complete analysis by \citet{DeckBatygin15} (see also \citealt{Batygin15}) 
for planets with comparable masses suggested that a significant portion
of the systems can be stably captured, and even for those exhibiting
overstability, the timescale the planets spend in/near resonance could be
comparable to the timescale the planets spend traveling between resonances.

\begin{figure}\centering
\includegraphics[width=\columnwidth]{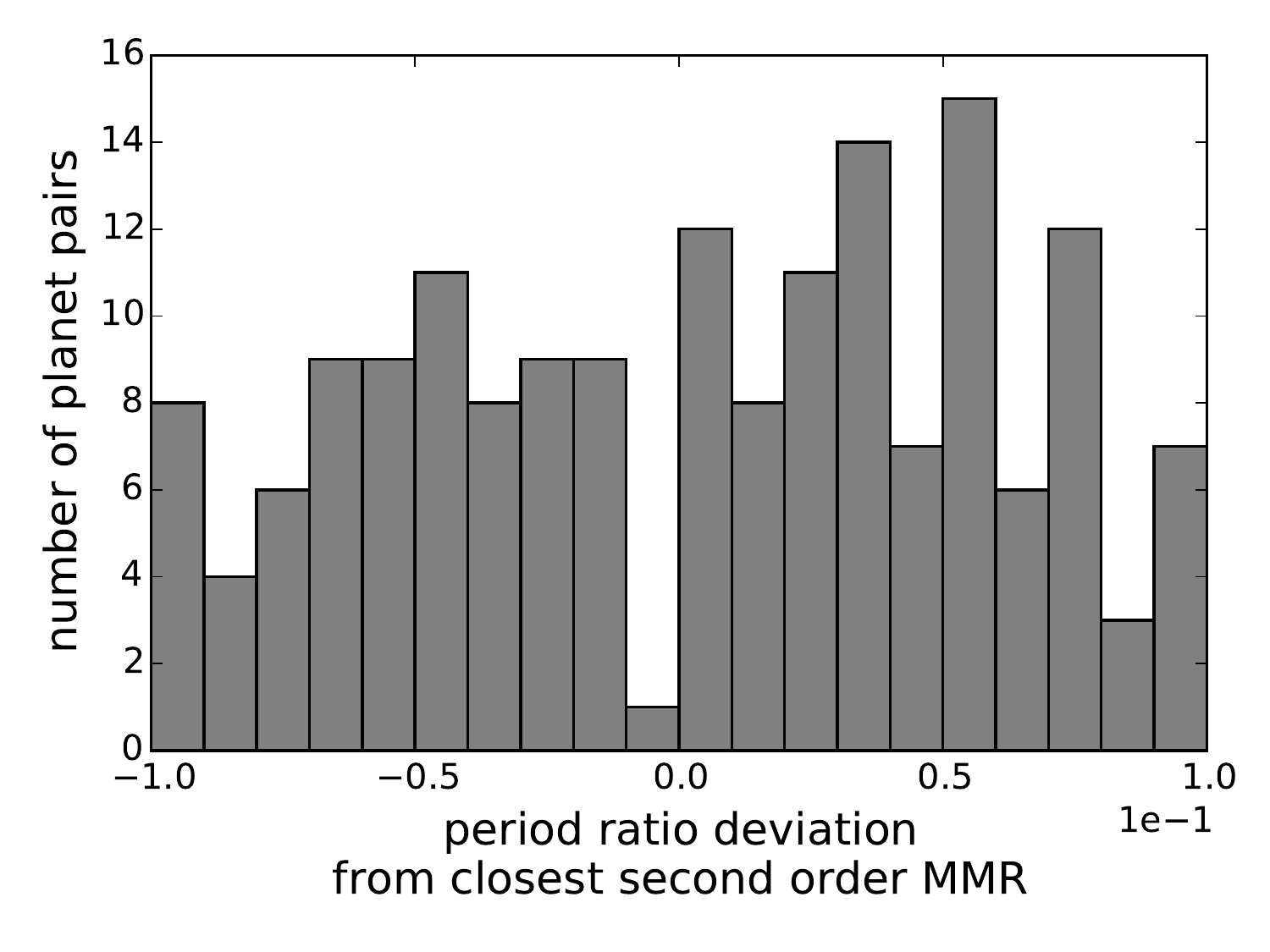}
\caption{Distribution of the period ratio deviation from the closest
  second-order MMR ($j:j-2$) with $j\leq 9$ for the observed multi-planet systems. Data
  from exoplanet.org is used. The deviation is given by 
  $P_2/P_1-j/(j-2)$, where $P_i$ are the periods of
  the planets. We see a paucity of planets just inside the resonance: For
  period ratio deviation in the range of $[-0.01,0]$, there is only one
  pair of planets, and this value is $3\sigma$ below the average.
}
\label{periodRatio}\end{figure}

While first-order MMRs for migrating planets have been studied in
depth, second-order ($j:j-2$) MMRs pose an equally compelling
problem. Among the observed multi-planet systems, there are a few pairs of 
planets that have period ratio very close to exact second-order resonances,
including Kepler 365 ($P_2/P_1-5/3=8.7\times 10^{-4}$),
Kepler 262 ($P_2/P_1-5/3=6.6\times 10^{-3}$),
Kepler 87 ($P_2/P_1-5/3=4.3\times 10^{-5}$),
Kepler 29 ($P_2/P_1-9/7=-4.4\times 10^{-5}$), and
Kepler 417 ($P_2/P_1-9/7=7.2\times 10^{-3}$).
It is unlikely that such a small deviation from commensurability is a
result of random chance; rather, it suggests that these planet pairs
are formed by resonance capture during migration.
In addition, the population statistics of multi-planet systems has
also revealed some signatures of second-order MMRs.  While the period
ratio distribution of {\it Kepler} multi's exhibits the most prominent
features near first-order MMRs (an excess at 3:2 and a deficit around
2:1; \citealt{Fabrycky14}) and at the enigmatic ratio of 2.2 \citep{SteffenHwang15},
the excess at the period ratio 1.7 ($\approx 5/3$) is also appreciable, 
and may be as significant as the 2:1 feature 
(see Fig.~20 of \citealt{SteffenHwang15}; J. Steffen, private communication).
The period ratio distribution near second-order MMRs is also worth noting: Fig.~\ref{periodRatio} 
shows that there is a paucity of planets right inside the resonances
[i.e. with period ratio close to, but smaller than, $j/(j-2)$].
Overall, these evidence suggests that second-order resonances are
strong enough to influence the architecture of multi-planet systems,
but they produce relatively few incidences of permanent capture.

{
Previous analytical studies on second and higher order MMRs are mostly in the context of solar system asteroids (e.g. \citealt{BorderiesGoldreich84,Lemaitre1984a,Lemaitre1984b}).
Recenty, \citet{Delisle15} studied analytically the resonance capture problem for arbitrary order MMRs with planets pairs having finite masses. Using an approximated integrable Hamiltonian \citep{Delisle14}, they derived a condition for stable capture in terms of the eccentricity damping timescales and equilibrium eccentricity ratio of the planets. They found that in order to make the captured MMR stable, the protoplanetray disk density profile often has to be locally inverted (i.e. the surface density increases outwards).
}
Other studies of second-order MMRs for planets migrating in
protoplanetary disks (e.g. \citealt{NelsonPapaloizou02,Xiang-GruessPapaloizou15}) have largely relied on numerical experiments.
While valuable, these numerical experiments are typically tailored toward
particular systems or setups, and it is often difficult to know
how the results depend on the physical inputs or parameters.  
There is also a common preconception that capture into second-order
MMRs is difficult because the planet perturbation associated with a
second-order resonance is too weak to counter the eccentricity damping
from planet interaction with the disk.  

In this paper and the companion paper, we develop an analytic theory
to study the capture and stability/escape of second-order MMRs for
planets pairs migrating in protoplanetary disks. {Unlike the first-order MMR, the Hamiltonian for the second-order MMR (for comparable mass planet pairs) is generally non-integrable. We treat this Hamiltonian exactly using a semi-analytic approach.} We also compare our
analytic results to numerical $N$-body experiments. Our goal is to derive the
conditions for resonance capture and the long-term stability of the
resonance, to understand the similarities and differences between
first and second-order MMRs, and to shed light on the observed
properties of MMRs in multi-planet systems.

Our paper is organized as follows. In Sections 2 and 3, we consider
second-order MMRs in the restricted three-body problem, where the
inner or outer planet is massless.  For such restricted problems, the
resonant Hamiltonian can be reduced to that of a one degree of freedom
system.  We use the reduced Hamiltonian to {review} the
resonance capture mechanism and the stability of the equilibrium
state.  In Section 4, we discuss the resonance capture criteria (in
terms of planet masses, migration and eccentricity damping rates), and
we examine whether planet migration in disks could allow resonance
capture and explain the observed \textit{Kepler} super-Earths in second-order
MMRs.  In Section 5, we advance to the general case when the two
planets have comparable masses, analyzing the capture mechanism and
finding the criterion for stable capture. {The stable capture criterion is compared to the result of \citet{Delisle15}.} Since the second-order
resonance Hamiltonian in this general case cannot be reduced to that
of a one degree of freedom system, the analysis of resonance capture
and stability is considerably more complicated than the restricted
problems, and we relegate much of the technical details to Appendix A (available online).
In Section 6, we perform N-body simulations of migrating planets and
compare with our analytic results derived in Sections 2-5. In Section
7 we summarize our key results and discuss their implications.

\section{Resonance in Restricted Three-Body Problem: Small Inner Planet}

Consider a system with two planets with mass $m_1, m_2$ orbiting
around a star with mass $M_\star$. Assume $a_1<a_2$, and let
$\alpha\equiv a_1/a_2<1$ be the semi-major axis ratio.
In this section we will consider the case when the inner planet's mass
is negligible ($m_1\ll m_2$). We assume $e_2=0$ and $e_1$ is small.
{Most of the results in this section has been covered in previous studies (e.g. \citealt{Delisle14,Delisle15}). Here we review this problem to prepare for the following discussion on capture criteria (Section 4) and the general problem with comparable mass planets (Section 5).}

\subsection{Hamiltonian}

At small eccentricity, the system's Hamiltonian $H$ near a $j:j-2$ MMR can be reduced to the following dimensionless form
\eq{\label{Ham_Inner}
-H = \eta\Theta+\Theta^2+\Theta\cos\theta,
}
where $\theta$ and $\Theta$ are a pair of conjugate coordinate and momentum, and $\eta$ is the ``resonance" parameter. They are given by
\eal{
&\theta = j\lambda_2+(2-j)\lambda_1-2\varpi_1,\\
&\Theta = \frac{3(j-2)^2}{16f_d \mu_2\alpha_0}\left(1-\sqrt{1-e_1^2}\right) \sim \mu_2^{-1}e_1^2,\\
&\label{eta_def}\eta \simeq \frac{1}{4f_d \mu_2} \left[(j-2)\alpha_0^{-1} -j\alpha_0^{1/2}\right],}
where $\mu_2=m_2/M_\star$.
The parameter $\alpha_0$ is related to $\alpha = a_1/a_2$ by
\eq{
\alpha_0\equiv \alpha[1+(j-2)e_1^2/2],
}
and is conserved because it is a function of a fast angle's conjugate momentum. Here $f_d$ is a function of $\alpha$ evaluated at $\alpha_0$, and is of order unity. \citet{MurrayDermott99} gives the expression of $f_d$ in Table 8.1, and the derivation of the above Hamiltonian and conserved quantities can also be found in that book\footnote{The scaling used in this paper is slightly different from that in \citet{MurrayDermott99}.}. In the dimensionless Hamiltonian, time is scaled to
\eq{
\tau = 4f_d\alpha_0\mu_2n_1t \equiv t/T_0,}
with
\eq{
T_0 = (4f_d\alpha_0\mu_2n_1)^{-1}.
}

\begin{figure*}\centering
\includegraphics[width=\textwidth]{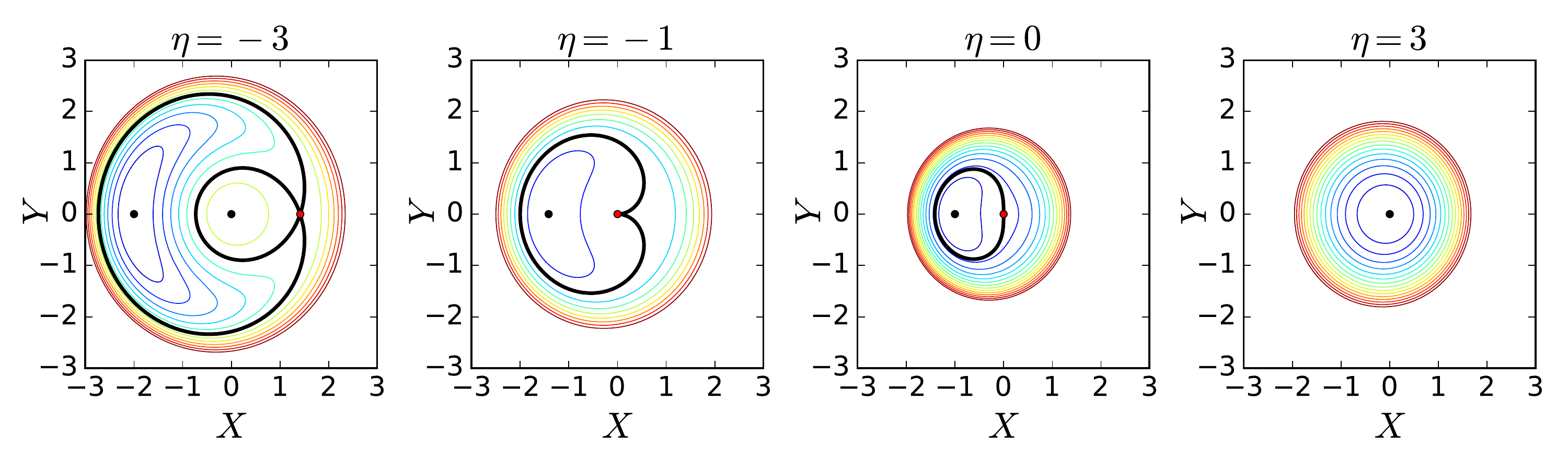}
\caption{Level curves of the Hamiltonian \eqref{Ham_Inner} for
  different $\eta$, plotted in the phase space of the conjugate
  variables $X=\sqrt{2\Theta}\cos\theta$ and
  $Y=\sqrt{2\Theta}\sin\theta$ (see Section 2.1). The thick black line in each panel
  marks the separatrix; the black (red) dots mark the stable
  (unstable) fixed points.}
\label{Hamiltonians}\end{figure*}

\begin{figure}\centering
\includegraphics[width=\columnwidth]{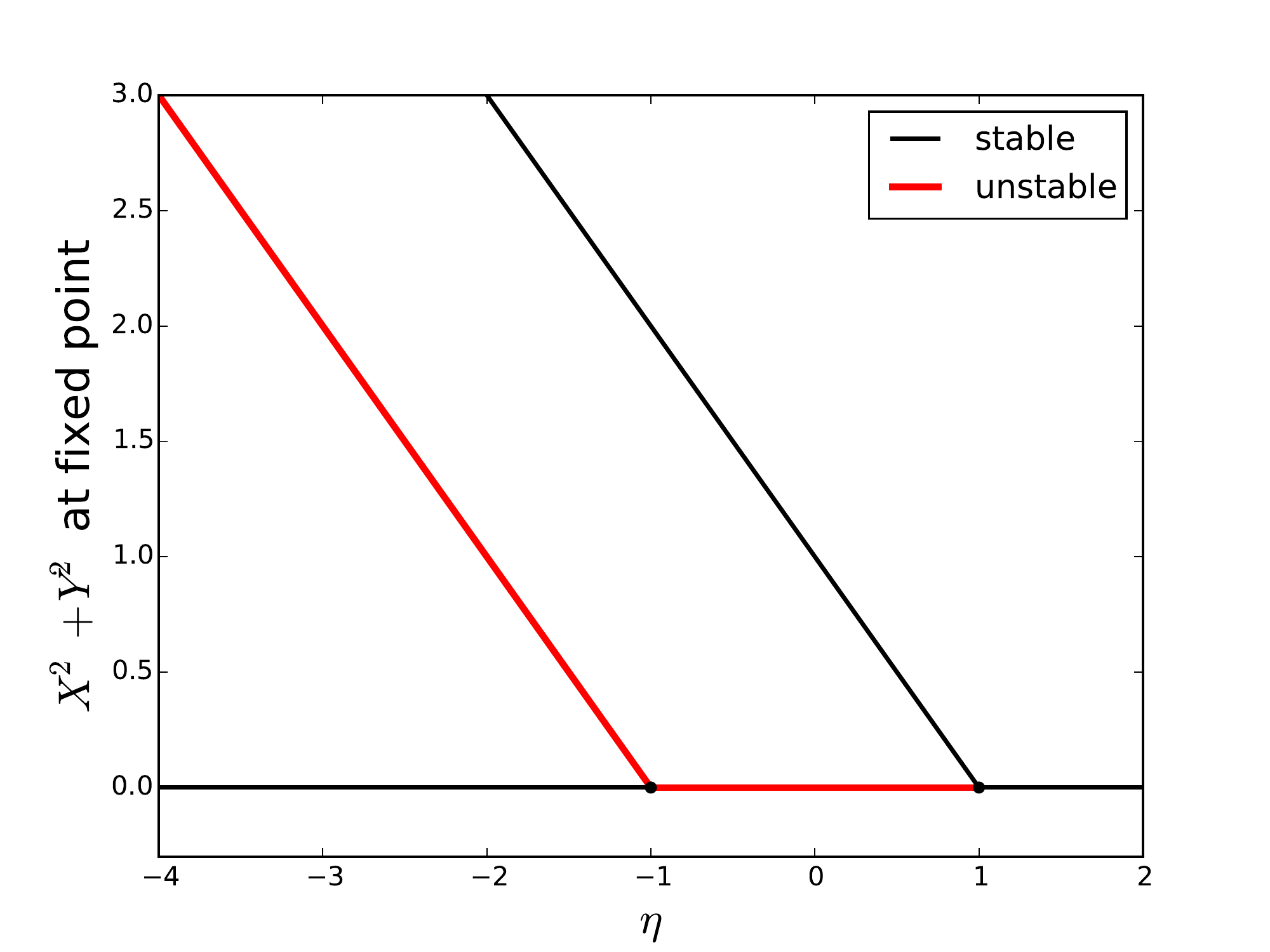}
\caption{Locations of the fixed points as a function of $\eta$ (see Section 2.1). The
  vertical axis shows the (squared) distance between the fixed point
  and the origin, $X^2+Y^2=2\Theta$, which is proportional to
  $e_1^2$. Bifurcations occur at $\eta=1$ and $\eta=-1$. The
  stability of the fixed points are marked by color; the red intervals are
  unstable.}
\label{fixedPtsLoc_toy}\end{figure}

The phase space topology of the Hamiltonian \eqref{Ham_Inner} gives
useful information about the structure of the resonance. Figure
\ref{Hamiltonians} shows the level curves of the Hamiltonian for
different $\eta$ values, plotted in the phase space of the conjugate
variables $X=\sqrt{2\Theta}\cos\theta$ and
$Y=\sqrt{2\Theta}\sin\theta$. Note that the distance to the origin is
$\sqrt{X^2+Y^2}=\sqrt{2\Theta}$, which is proportional to $e_1$. There
can be at most three fixed points, located at $Y=0$ and $X=0$,
$X=-\sqrt{1-\eta}$ (for $\eta<1$) and $X=\sqrt{-1-\eta}$ (for
$\eta<-1$), and their distances to the origin as a function of $\eta$ are
shown in Figure \ref{fixedPtsLoc_toy}. Two bifurcations take place as
$\eta$ increases from $\eta\ll -1$: at $\eta=-1$ the unstable fixed
point ($X=\sqrt{-1-\eta}$) merges with the origin, making the origin
unstable; at $\eta=1$ the stable fixed point ($X=-\sqrt{1-\eta}$)
merges with the origin (which is now unstable), making the origin
stable again. For $\eta<-1$ or $\eta>1$, the origin is a stable fixed
point and orbit with low eccentricity stays at low eccentricity. For
$-1<\eta<1$, on the other hand, planets with initially near circular
orbit will exhibit oscillating eccentricity, with a maximum
$\Theta\sim 1-\eta$ [corresponding to
  $e_1\sim\mu_2^{1/2}(1-\eta)^{1/2}]$. The width of the resonance can
be defined as the range of $\alpha_0$ for which $-1<\eta<1$; since
$|\partial\eta/\partial\alpha_0|\sim\mu_2^{-1}$, the width in terms of
$\alpha_0$ is $\sim\mu_2$. This width of resonance is small, which
means that few planets would be in resonance if they are formed
completely in situ.

\subsection{Eccentricity excitation and resonance capture}

The Hamiltonian \eqref{Ham_Inner} only includes the non-dissipative
resonant interaction between the two planets. In real systems, the
planets can experience (dissipative) perturbations from other sources,
e.g. planet-disk interaction. When the resonant interaction between
the planets is dominant (i.e. the characteristic timescales of other
perturbations are all $\gg T_0$), the effects of the other
perturbations can be included as a slow variation of $\eta$, with
$|d\eta/d\tau|\ll 1$. Therefore, it is helpful to first study how the
system evolves when it passes the resonance as $\eta$ slowly
increases/decreases [see, e.g. \citet{BorderiesGoldreich84, XuLai16}]. We assume that the initial (out-of-resonance) eccentricity
of the planet (test mass) is sufficiently small such that $\Theta_0\ll
1$, or equivalently $e_0\ll \mu_2^{1/2}$. The two possibilities are

(i) When the system passes resonance from $\eta<-1$ (i.e. $\eta$ is
slowly increasing), the planet's eccentricity is excited at $\eta =
-1$ because the origin ($\Theta=0$) becomes unstable. After that, the
phase space area bounded by the trajectory is conserved because the
evolution is adiabatic. This area is of order unity, therefore the
system ends up with a final eccentricity $\Theta_f\sim 1$ or
$e_f\sim\sqrt{\mu_2}$. Note that the system is not in resonance, since
the resonant angle $\theta$ is circulating.

(ii) On the other hand, when the system passes resonance from $\eta>1$
(i.e. $\eta$ slowly decreasing), the planet can be captured into
resonance. This is because as $\eta$ goes below $1$, the origin
becomes unstable and the stable fixed point moves away from the
origin. When $|d\eta/d\tau|\ll 1$, the system follows this stable
fixed point (with $2\Theta=1-\eta$) and advects into the libration
zone with a finite eccentricity. The eccentricity excited by this
resonant advection is unbounded as long as $\eta$ keeps decreasing and
the small eccentricity approximation holds.

\subsection{Effect of planet migration}

We now analyze the effect of planet migration due to interaction with
a protoplanetary disk. The dissipative effect of planet-disk interaction can be
parameterized by
\eal{
\label{diss_1}&\left.\frac{1}{a_i}\frac{da_i}{dt}\right|_{\rm diss} = -\frac{1}{T_{m,i}}-\frac{p_ie_i^2}{T_{e,i}},\\
\label{diss_2}&\left.\frac{1}{e_i}\frac{de_i}{dt}\right|_{\rm diss} = -\frac{1}{T_{e,i}},
}
where $T_{m,i}$ and $T_{e,i}$ (with $i=1,2$) characterize the timescale of inward migration and eccentricity damping; and we define the dimensionless timescales $\tau_{m,i}\equiv T_{e,i}/T_0,\tau_{e,i}\equiv T_{e,i}/T_0$. For small (non-gap-opening) planets in a gaseous disk, these timescales are given by (e.g. \citealt{Ward97,GoldreichSchlichting14})
\eal{
\label{Tm}&T_{m,i}^{-1}\sim \mu_i\mu_{d,i}\left(\frac{a_i}{h_i}\right)^2n_i,\\
\label{Te}&T_{e,i}^{-1}\sim \mu_i\mu_{d,i}\left(\frac{a_i}{h_i}\right)^4n_i,
}
where $\mu_{d,i}=\Sigma_da_i^2/M_\star$ is disk to star mass ratio and
$h_i$ is the disk's scale hight at $a_i$. For an Earth-mass planet at
$a_i\sim 0.5$AU, with $\mu_{d,i}\sim 10^{-4}$ and $h_i/a_i\sim 0.1$,
the migration time $T_{m.i}$ is of order 1Myr, while $T_{e,i}\sim
10^4$yrs. The actual value of $T_{m,i}$ and $T_{e,i}$ are uncertain
since they depend on the thermodynamic property and profile of the
disk (e.g. \citealt{Baruteau14}). The parameter $p_i$ characterizes
the energy dissipation rate associated with eccentricity damping, and
usually $p_i\simeq2$.

Noting that $\partial\eta/\partial\alpha_0<0$,
$d\ln\alpha_0/d\tau=d\ln\alpha/d\tau+(j-2)e_1 (d e_1/d\tau)$,
and using Eqs.~\eqref{diss_1} - \eqref{diss_2} we find
\eq{\label{deta_dtau}
\frac{d\eta}{d\tau} = \left|\frac{\partial\eta}{\partial\ln\alpha_0}\right|\left[-\frac{1}{\tau_m}+\frac{(p_1+j-2)e_1^2}{\tau_{e}}\right],
}
where $\tau_e\equiv\tau_{e,1}$ and
\eq{\label{taum_def}
\frac{1}{\tau_m}\equiv -\frac{1}{\tau_{m,1}}+\frac{1}{\tau_{m,2}}.
}
Note that $\tau_e,\tau_m$ correspond to the physical timescales
$T_e=T_0\tau_e$ and $T_m=T_0\tau_m$. To capture the system in
resonance, we require $T_m>0$, i.e. $T_{m,1}>T_{m,2}$. We will focus
on this (convergent migration) case in the remainder of this paper.

From Eq. \eqref{deta_dtau} we see that when the eccentricity grows to a certain value, the second term in $d\eta/d\tau$ balances the first term and the system enters an equilibrium state. Therefore, the eccentricity will not grow unbounded and the system can be trapped in the resonance with a fixed $\eta$ (corresponding to a fixed period ratio) for a long time. In this equilibrium state,
\eq{
e_{1,\rm eq} = \sqrt{\frac{\tau_e}{(p_1+j-2)\tau_m}}.
}
\subsection{Stability of capture}
Planet migration and eccentricity damping due to planet-disk
interaction are responsible for resonance capture and the
establishment of the equilibrium; they also affect the stability of
the equilibrium.

\begin{figure}\centering
\includegraphics[width=\columnwidth]{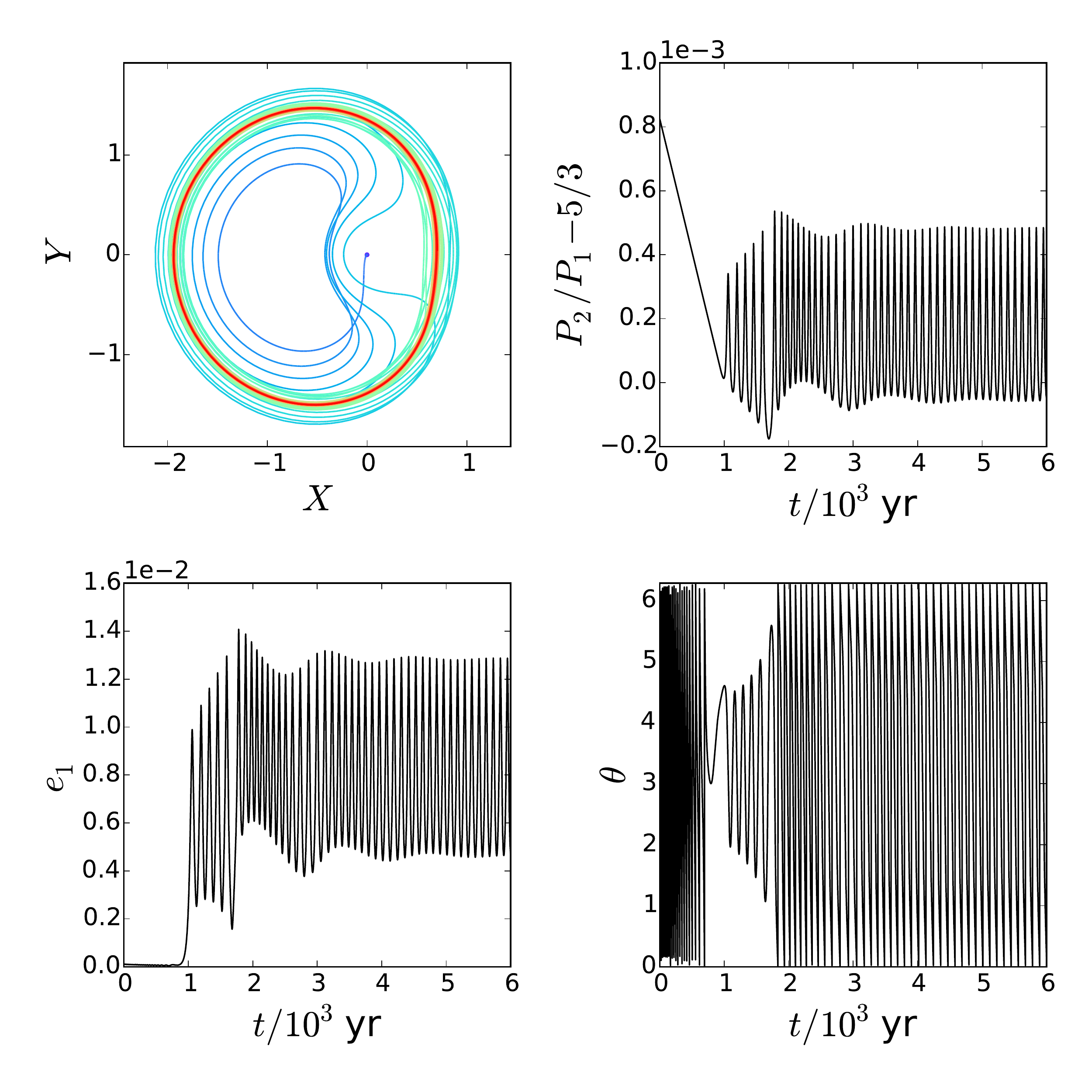}
\caption[Caption for LOF]{Evolution of a system near 3:5 mean motion
  resonance with $M_\star=1M_\odot,m_2=10M_\oplus, m_1\ll m_2$, and
  $a_1=0.1$AU (see Section 2.4). The migration and eccentricity damping timescales are
  $T_m=3$Myr and $T_e=1$kyr, corresponding to $\eta_{\rm
    eq}=-0.52$. The upper left panel shows the phase space trajectory,
  with time marked by color (blue indicates earlier time, followed by
  green, and red indicates later time); the upper right panel shows
  the deviation of period ratio $P_2/P_1$ from the exact resonance; the
  lower panels show the evolution of eccentricity and resonant angle
  $\theta$. We see that as the system is captured into resonance (with
  $\theta$ librating) at $t\sim 1$kyr. However, the phase space area
  bounded by the trajectory and the libration amplitude of $\theta$
  increase due to overstability. This eventually pushes the system out
  of resonance ($\theta$ begins circulating at $t\simeq 2$kyr). The
  system ends up in an quasi-equilibrium state with the period ratio
  and eccentricity oscillating at fixed amplitudes.\footnotemark}
\label{toyModelOverstb2}\end{figure}
\begin{figure}\centering
\includegraphics[width=\columnwidth]{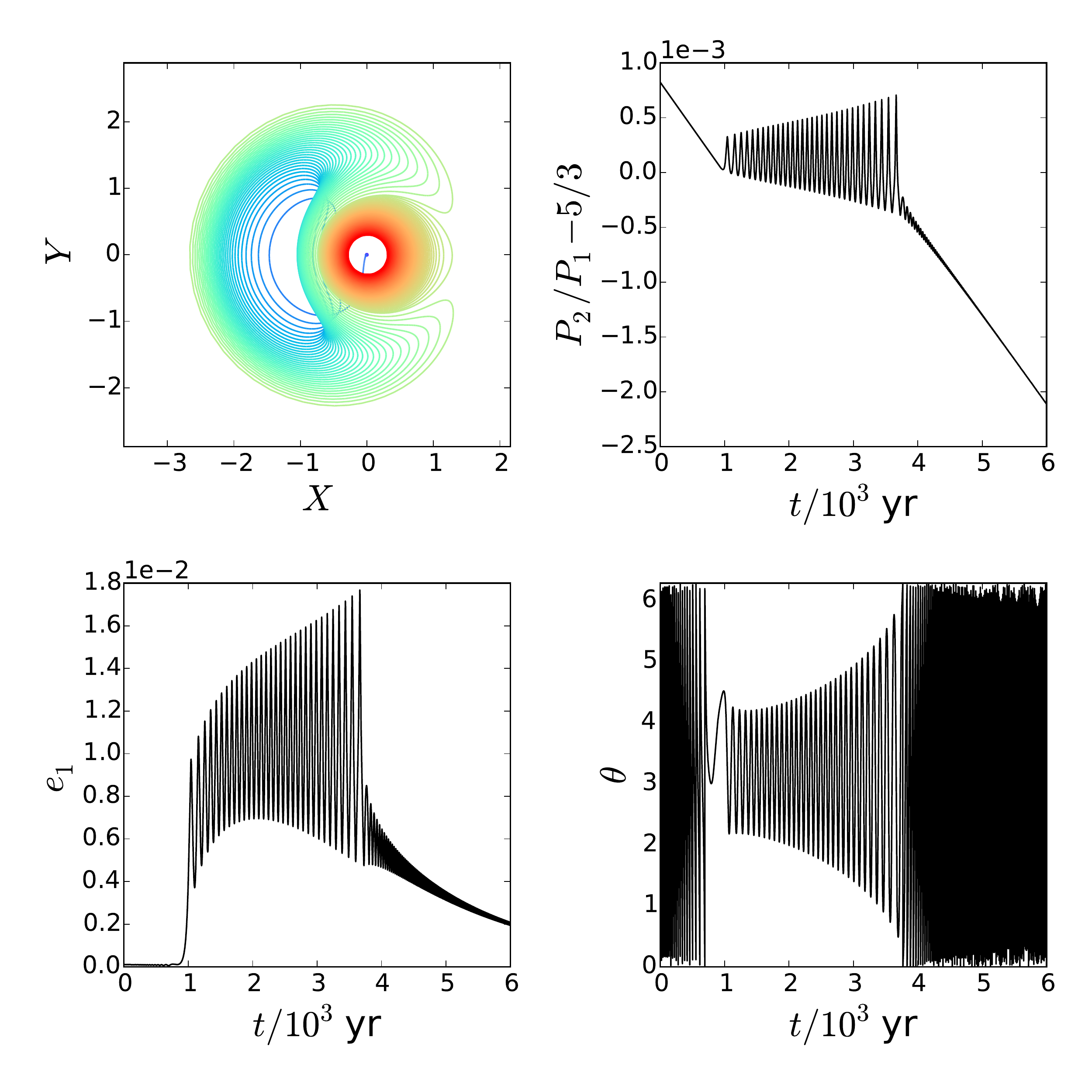}
\caption{Same as Figure \ref{toyModelOverstb2}, except that $T_e$ is
  increased to 2kyr, giving $\eta_{\rm eq}=-1.27$ (see Section 2.4). In this case,
  the system enters the inner circulating zone upon exiting the
  libration zone. Eventually, the period ratio is no longer held near
  the equilibrium, and the eccentricity excited in resonance is quickly
  damped due to planet-disk interaction.}
\label{toyModelOverstb1}\end{figure}

\footnotetext{The choice of $T_m$ and $T_e$ in Figure
  \ref{toyModelOverstb2}, \ref{toyModelOverstb1} and \ref{toyModelStb}
  gives a very low $T_e/T_m$ which is unrealistic in most systems;
  here we are choosing these values in order to better illustrate the
  capture process (for realistic parameters, the trajectory varies too
  little each cycle). For systems in Figure \ref{toyModelOverstb1} and
  \ref{toyModelStb}, changing $T_e$ to more realistic values
  (e.g. $T_e\sim10^{-2}T_m$) doesn't affect the result. For Figure
  \ref{toyModelOverstb2}, however, the behavior of the system is due
  to the small $T_e/T_m$, suggesting that this case is
  unlikely to happen for more realistic $T_e/T_m\sim 10^{-2}$ 
  (see text for discussion).}

The equations of motion of the system near the MMR is determined by the 
Hamiltonian (\ref{Ham_Inner}). Including the eccentricity damping term
due to planet-disk interaction, we have
\eal{
\label{eom_Th}& \dot\Theta = -\Theta\sin\theta - \frac{2}{\tau_{e}}\Theta,\\
\label{eom_th}& \dot\theta = -\eta-2\Theta-\cos\theta,}
where $\dot\Theta=d\Theta/d\tau$ and $\dot\theta=d\theta/d\tau$. In
addition, the parameter $\eta$ evolves according to
Eq.~\eqref{deta_dtau}, or
\eq{
\dot\eta = -\frac{\beta}{\tau_{m}} + \frac{4(p_1+j-2)}{(j-2)\tau_e}\Theta,
\label{eom_eta}}
where
\eq{
\beta \equiv \left|\frac{\partial \eta}{\partial\ln \alpha_0}\right|
\simeq \frac{3}{8f_d \mu_2} j^{2/3}(j-2)^{1/3}.}
Note that $\Theta=[(j-2)\beta/4]e^2$. For slow migration, $\beta/\tau_m$ and $1/\tau_e$ are both very small. The equilibrium point, to lowest order, is given by
\eal{
&\Theta_{\rm eq} = \frac{(j-2)\beta \tau_{e}}{4(p_1+j-2)\tau_{m}},\\
&\theta_{\rm eq} \simeq \pi+\frac{2}{\tau_{e}},\\
&\eta_{\rm eq} \simeq 1-2\Theta_{\rm eq}.\label{eq:etaeq1}
}
The linearized equations are (with $\Theta = \Theta_{\rm eq}+\delta\Theta$ and so on)
\eq{
&\delta\dot\Theta = \Theta_{\rm eq}\delta\theta,\\
&\delta\dot\theta = -\delta\eta - 2\delta\Theta - \frac{2}{\tau_{e}}\delta\theta,\\
&\delta\dot\eta = \frac{4(p_1+j-2)}{(j-2)\tau_{e}}\delta\Theta,
}
and the characteristic equation is (for $\delta\Theta,\delta\theta,\delta\eta\propto e^{\lambda t}$)
\eq{\label{char_eqn}
\lambda^3 + \frac{2}{\tau_{e}}\lambda^2 + 2\Theta_{\rm eq}\lambda + \frac{\beta}{\tau_m}=0.
}
It is easy to see that the system has one negative real eigenvalue and two complex eigenvalues. We can assume $\Theta_{\rm eq}\gtrsim 1$ (corresponding to $\eta_{\rm eq}\lesssim -1$; otherwise the system can always stay captured; see below and Figure \ref{toyModelOverstb2}), and we notice $\tau_e^{-1},\beta/\tau_m\sim \Theta_{\rm eq}/\tau_e\ll 1$; therefore the complex eigenvalues are approximately
\eq{
\lambda \simeq \lambda_r\pm i\sqrt{2\Theta_{\rm eq}},
}
where the real part $\lambda_r$ is small ($|\lambda_r|\sim \tau_e^{-1}$). Substituting this back to \eqref{char_eqn} gives 
\eq{
\lambda_r\simeq \frac{p_1}{(j-2)\tau_e}.
}
This indicates that the oscillation around the equilibrium is overstable for all realistic ($p_1>0$) situations. {This agrees with the result of \citet{Delisle15}.}

When the oscillation near the equilibrium is overstable, the phase space area bounded by the trajectory will eventually exceed the area of the libration zone and the system will be pushed out of resonance. Whether the system can maintain an approximately constant period ratio depends on the value of $\eta$ at the equilibrium state.

Figure \ref{toyModelOverstb2} shows an example when $\eta_{\rm eq}>-1$. In this case, after resonance capture, the system is pushed into the outer circulation zone due to overstability. However, $\eta$ is still held near the equilibrium value because the eccentricity excitation forced by the unstable origin can balance the eccentricity damping. Therefore, $\alpha$ (and the period ratio) is still held approximately constant despite the fact that the system, strictly speaking, is not in resonance. Note that the condition $\eta_{\rm eq}>-1$ requires
\eq{
\Theta_{\rm eq}=\frac{3}{32}\frac{j^{2/3}(j-2)^{4/3}}{(p_1+j-2)f_d}\mu_2^{-1}\frac{T_e}{T_m}<1,
}
which corresponds to
\eq{
\frac{T_e}{T_m}\lesssim \mu_2
}
Since $T_e/T_m\sim(h/a)^2\sim 10^{-3}$ - $10^{-2}$, this condition can be satisfied only for massive planets ($\mu_2\gtrsim 10^{-3}$ - $10^{-2}$).

When $\eta_{\rm eq}<-1$, however, the system will eventually\footnote{For some cases, the system might first stay in the outer circulating zone for some time, but it usually lasts for only a few cycles because this state is also unstable and the system tends to cross the separatrix to enter the inner circulating zone.} enter the circulation zone centered at the origin, as is shown in Figure \ref{toyModelOverstb1}. Within this zone, there is no eccentricity excitation mechanism and the eccentricity decreases. As a result, the system's eccentricity can no longer hold $\eta$ near the equilibrium, and the period ratio increasingly deviates from the resonant value. In this case, the system stays near resonance for a duration of order
\eq{
T_0|\lambda_r|^{-1}\sim T_0\tau_e = T_e.
}
Since we know that $T_e\ll T_m$, while $T_m$ is the timescale for migrating between resonances, the system should be out of resonance for most of the time.

\section{Resonance in Restricted Three-Body Problem: Small Outer Planet}

Next we consider the case when the inner planet is more massive and
the outer planet has negligible mass, i.e. $m_1\gg m_2$. {This case can
be illuminating since previous studies \citep{Delisle15, DeckBatygin15} already showed that the behavior of the system,
especially the stability of capture, may depend on the
mass ratio.}

Similar to Section 2, the Hamiltonian can be written in the form of Eq. \eqref{Ham_Inner}, but with the time scaled to
\eq{
\tau= 4f_d \mu_1 n_2t\equiv t/T_0,
}
where $\mu_1=m_1/M_\star$. The conjugate variables and the ``resonance" parameter are
\eal{
&\theta = j\lambda_2+(2-j)\lambda_1-2\varpi_2,\\
&\Theta = \frac{3j^2}{16f_d\mu_1} \left(1-\sqrt{1-e_2^2}\right),\\
&\eta\simeq \frac{1}{4f_d \mu_1} \left[(j-2)\alpha_0^{-3/2}-j\right],
}
where $\mu_1=m_1/M_\star$, and
\eq{
\alpha_0\equiv \alpha(1+je_2^2/2)
} 
is conserved.


With $\partial\eta/\partial\alpha_0<0$, we have
\eq{
\frac{d\eta}{d\tau} = \left|\frac{\partial\eta}{\partial\ln\alpha_0}\right|\left[-\frac{1}{\tau_m}+\frac{(j-p_2)e_2^2}{\tau_e}\right],
}
where $\tau_m$ is the same as in Eq. \eqref{taum_def} while $\tau_e\equiv\tau_{e,2}$. Resonance capture still occurs when the outer planet migrates inward more quickly (i.e. convergent migration), and the equilibrium eccentricity is
\eq{
e_{2,\rm eq} = \sqrt{\frac{\tau_e}{(j-p_2)\tau_m}}.
}

\begin{figure}\centering
\includegraphics[width=\columnwidth]{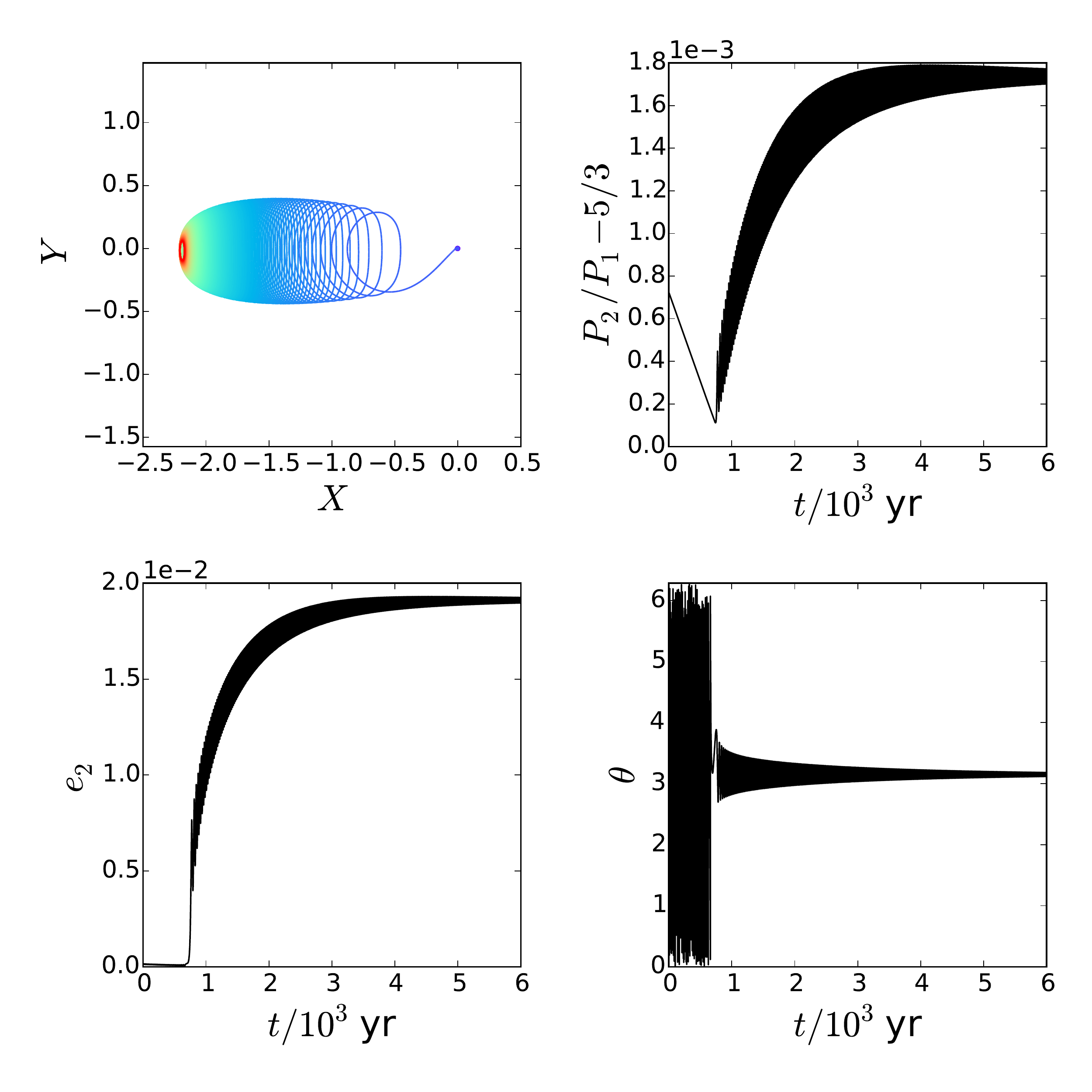}
\caption{Evolution of a system near 3:5 mean motion resonance with
  $M_\star=1M_\odot,m_1=10M_\oplus, m_2\ll m_1$, and $a_1=0.1$AU (see Section 3.3). The
  migration and eccentricity damping timescales are $T_m=3$Myr and
  $T_e=1$kyr. We see that the system quickly approaches the
  equilibrium state, and the libration amplitude of $\theta, e$ and
  period ratio decrease. In the end, the system's trajectory converges
  to the fixed point, and $\theta=\pi$.}
\label{toyModelStb}\end{figure}

The equations of motion are still given by Eqs. \eqref{eom_Th} -
\eqref{eom_th}, while \eqref{eom_eta} is replaced by
\eq{\label{eom_eta_2}
\dot\eta = -\frac{\beta}{\tau_{m}} + \frac{4(j-p_2)}{j\tau_e}\Theta,
}
where
\eq{
\beta \equiv \left|\frac{\partial \eta}{\partial\ln \alpha_0}\right|
\simeq \frac{3 j}{8f_d\mu_1}.}
{The equilibrium eccentricity is given by
\eal{
&\Theta_{\rm eq} = \frac{j\beta \tau_{e}}{4(j-p_2)\tau_{m}},
}
and the characteristic equation is}
\eq{
\lambda^3 + \frac{2}{\tau_{e}}\lambda^2 + \frac{j\beta \tau_{e}}{2(j-p_2)\tau_{m}}\lambda + \frac{\beta}{\tau_m}=0.
}
Solving this equation shows that the eigenvalues all have negative real part as long as $p_2>0$. Therefore, the equilibrium state is stable, and the trajectory converges to the fixed point, giving a fixed resonant angle $\theta=\pi$. {This agrees with the result of \citet{Delisle15}.}

Thus, in contrast to the $m_1\ll m_2$ case considered in Section 2, convergent migration in the $m_1\gg m_2$ case leads to permanent capture into the resonance. Figure \ref{toyModelStb} shows an example of such permanent capture, where the eccentricity, period ratio and resonant angle all converge to fixed values.

\section{Conditions for Resonant Capture During Migration}

We know from the previous sections that a pair of planets can be
captured into resonance (either temporarily or permanently) if they
undergo convergent migration with sufficiently large $T_e$ and
$T_m$. On the other hand, it is evident that when $T_e$ or $T_m$ is
too small, the system cannot be captured: for small $T_e$ the
excitation of eccentricity due to resonant motion is fully suppressed,
while for small $T_m$ the system passes the resonance so fast that it
has no time to excite the eccentricity to large enough value before
the origin ($e=0$) becomes stable again. In this section, we obtain
the criteria that $T_e$ and $T_m$ must satisfy in order to allow resonance capture.

First we consider the constraint on $T_e$. Resonance capture requires
that the eccentricity damping due to planet-disk interaction be weaker than
the eccentricity excitation due to resonant interaction, i.e.,
\eq{
T_e\gtrsim T_0, \text{~~or~~}\tau_e\gtrsim 1.
}

The constraint on $T_m$ is slightly more complicated. A successful
capture during convergent migration (decreasing $\eta$) requires that
the system have plenty of time to catch up to the stable fixed point at
$\Theta=(1-\eta)/2$ before the origin ($\Theta=0$) becomes stable
again (see Figs. \ref{Hamiltonians} - \ref{fixedPtsLoc_toy}). In other
words, $|\dot\Theta|$ should exceed $|\dot\eta|$ before $\eta$ reaches
-1. From the equation of motion [see Eq.~\eqref{eom_Th}] we see that
$|\dot\Theta|\sim \Theta$; thus we require
\eq{
\dot\Theta(\eta=-1)\sim \Theta_0 e^{2/|\dot\eta|} \gtrsim |\dot\eta|
}
where $\Theta_0$ is the value of $\Theta$ when the system enters the resonance (at $\eta=1$). Since $|\dot\eta|\sim\beta/\tau_m\sim 1/(\mu\tau_m)$, where $\mu\equiv \mu_1+\mu_2$ [see Eq. \eqref{eom_eta} or \eqref{eom_eta_2}], the above condition becomes
\eq{
\mu\tau_m\gtrsim -\ln\Theta_0.
} 
It is worth noting that when $\mu\tau_m\sim-\ln\Theta_0$, whether
resonance capture is successful also depends on the initial value of
the resonant angle $\theta$. Since the initial $\theta$ is essentially arbitrary, capture
can be considered as probabilistic in this case. The probabilistic
capture regime only occupies a relatively small portion of the
parameter space when $\Theta_0\lesssim 1$ (Xu \& Lai 2016).

In the above calculation we have assumed $\Theta_0\lesssim 1$. For
$\Theta_0\gtrsim 1$, \citet{BorderiesGoldreich84} showed that
resonance capture becomes probabilistic for infinitely large $T_e$ and
$T_m$; and for large $\Theta_0$ (e.g. $\Theta_0\gtrsim 10$) the
probability of capture is negligible. For smaller $T_e$ and
$T_m$, we expect that the capture probability can only be
lower. Therefore, for $\Theta_0\gtrsim 1$, resonance capture is unlikely.

In summary, the conditions for resonance capture are
\eq{
\tau_e\gtrsim 1,~~~\mu\tau_m\gtrsim -\ln\Theta_0,~~~\Theta_0\lesssim 1.
}
In terms of physical quantities, these conditions are
\eq{\label{trap_cond_1}
T_e\gtrsim T_0\sim \frac{P_1}{8\pi\mu},~~~T_m\gtrsim \frac{P_1}{8\pi\mu^2}\ln\frac{\mu}{e_0^2},~~~e_0\lesssim \mu^{1/2}
}
where $P_1$ is the inner planet's period, and $e_0$ is the
``initial'' eccentricity when the planet enters resonance.

For planets migrating in a disk, the initial eccentricity $e_0$ is
usually small so the last condition in (\ref{trap_cond_1}) is
satisfied. Due to the non-resonant perturbation from other planets, the
initial eccentricity is at least of order $e_0\sim \mu$, which
yields $(-\ln\Theta_0)\sim \ln\mu$. Thus resonance capture
requires 
\eq{ T_e\gtrsim \frac{P_1}{8\pi\mu},~~~T_m\gtrsim
  \frac{5P_1}{4\pi\mu^2}\left|\frac{\ln\mu}{\ln 10^{-5}}\right|.  
\label{eq:criteria}}
Here we have used $\ln 10^{-5}\simeq -10$. For typical protoplanetary
disks, $T_e/T_m\sim (h/a)^2$ is about $10^{-2}$ to $10^{-3}$. This implies that
for most planets (with $\mu\lesssim T_e/T_m$), and the condition for $T_e$ is less
demanding and does not need to be considered.

\begin{figure}
\centering
\includegraphics[width=\columnwidth]{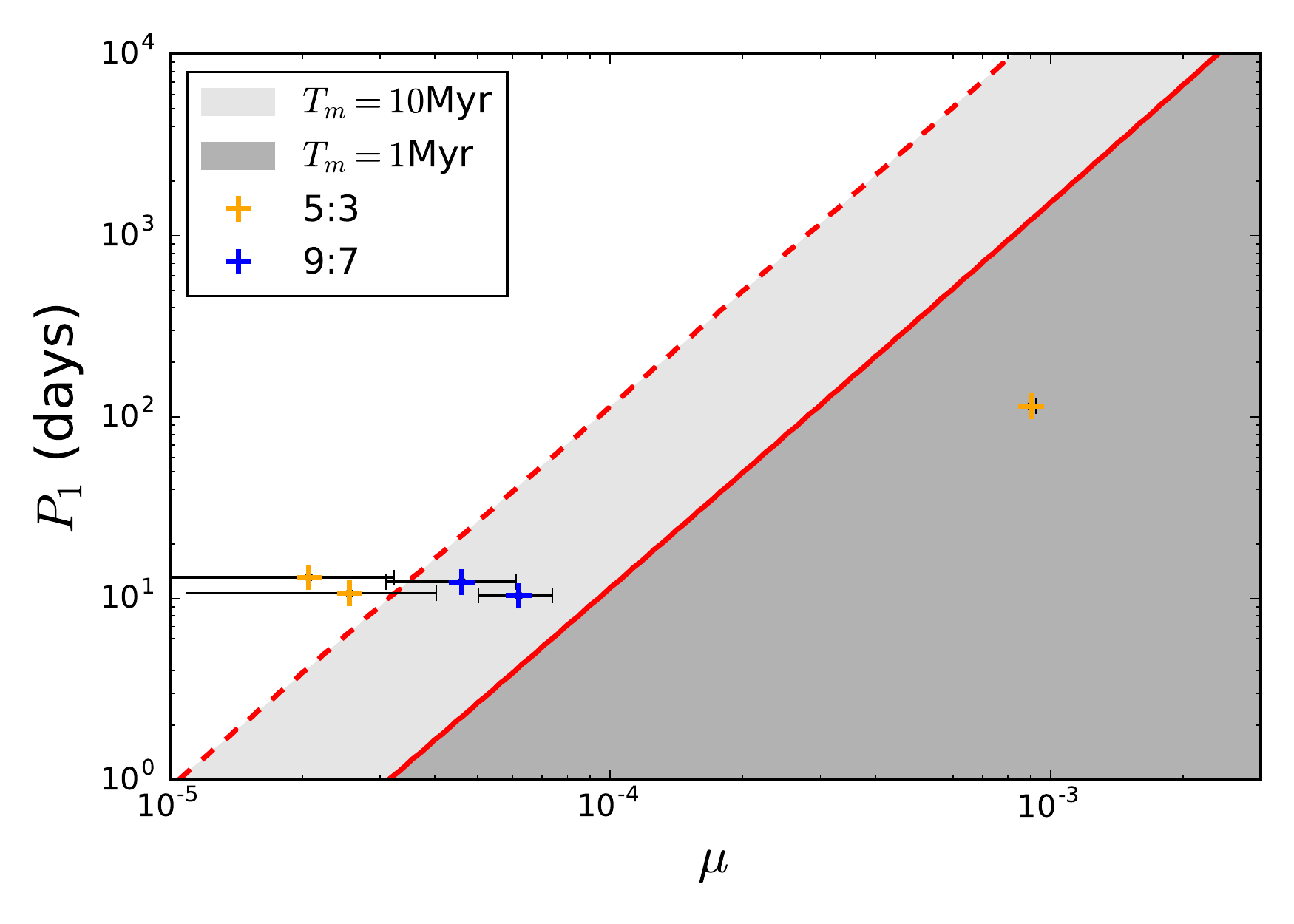}
\caption{Estimation for the region in the $(\mu,P_1)$ parameter space
  where resonance capture is possible. Here $\mu=\mu_1+\mu_2=(m_1+m_2)/M_\star$
and $P_1$ is the orbital period of the inner planet.
In the dark (light) grey regions, the system can be captured for $T_m=1$~Myr ($10$~Myr),
with the solid (dashed) line showing the boundary [the second
inequality of (\ref{eq:criteria})].
The crosses mark the five known systems that are close to
(i.e. with period ratio deviation less than $6\times 10^{-3}$)
the second-order MMR;
the color of the marker labels different MMR, as indicated, \textbf{and the error bars mark the uncertainty of $\mu$}. 
All these systems lie inside or close to the estimated resonance capture region.}
\label{trapRegion}
\end{figure}

Figure \ref{trapRegion} shows the region in the $(\mu, P_1)$ parameter
space where resonance capture is allowed for $T_m=1$~Myr and 10~Myr. We
see that there is a significant region in the parameter space where
resonance trapping is possible, especially for more massive planets
($\mu\gtrsim 10^{-3}$). In Figure \ref{trapRegion} we also compare the
observed planet pairs in second-order MMR with our estimation; we
see that these observed planet pairs lie within or close to our
estimated boundary, suggesting that the capture mechanism being
studied may account for their formation.
Some of the planets appear to be outside the depicted capture region;
these planets may require somewhat larger $T_m$ ($\gtrsim 10$~Myr) to
allow capture (e.g., as when the planet migrates in low-density disks or when $T_{m,1}\simeq T_{m,2}$).
Note that our resonance capture criteria are
estimates, so the actual boundaries can be easily shifted by a factor
of a few compared to those shown in Fig.~\ref{trapRegion}.

It is worth noting that among the five systems close to the MMR there
is only one system with massive planet ($\mu\gtrsim 10^{-3}$),
although from \eqref{Tm} and \eqref{eq:criteria} we see that capture is easier when $\mu$ is greater. 
This may reflect the fact that giant planets are less common than low-mass planets
(super-Earths and sub-neptunes);
other possible reasons for this disagreement are discussed in Section 7.2.

\section{Resonance of Two Planets of Comparable Masses}

There are two motivations to study second-order MMR involving two
similar mass planets. First, our analysis of the restricted three-body
problem (Sections 2 and 3) shows that the stability property of the
equilibrium following resonance capture is completely different in the
two limiting cases ($m_1\ll m_2$ and $m_2\ll m_1$).  It is necessary
to know where the transition between the two behaviors (stable vs
overstable equilibrium state) occurs, and whether the values of $T_e$
and $T_m$ affect the stability. Second, four of the five observed
planet pairs close to second-order resonances have mass ratio close to
unity, so a model for similar mass planets in resonance is necessary
if we want to compare our theoretical result with observations.

{
Previous studies \citep{Delisle14, Delisle15} calculate the stability of the captured state for two planets with general mass ratio using an integrable Hamiltonian. This Hamiltonian is exact in the two limiting cases ($m_1\ll m_2$ and $m_2\ll m_1$), but is approximate for planets with comparable masses. 
}
In this section we first derive the {full non-integrable} resonance Hamiltonian for the general
($m_1\sim m_2$) case, and study the mechanism of resonance capture by analyzing
the fixed points of the system. Then we obtain the conditions for
resonance capture, and compare these with our results for the
limiting cases. Finally we calculate how the stability and escape time
(i.e. time inside resonance before escaping, when the capture is
unstable) depends on $T_m$ and $T_e$, {and compare the results with \citet{Delisle15}}. Due to the complexity of
algebra, we omit most of the calculation in the main text, and
relegate the details to Appendix A (available online).

\subsection{Hamiltonian and fixed points}

For first-order resonances, the Hamiltonian of planets with comparable
masses can be transformed to that of a one degree of freedom system,
of the same form as the restricted problem \citep{Sessin84,Wisdom86,Henrard86,Deck13}. 
However, such simplification is impossible for second-order
resonances. This is because the leading terms of the resonant and secular
perturbations have similar strengths, and the ``mixing" of the two
perturbations obstruct simplification. {\citet{Delisle14} propose a method to reduce the Hamiltonian of an arbitrary order MMR to an integrable form which has the same mathematical form as the Hamiltonian in the limiting cases, but this reduction involves nontrivial approximation for second and higher order MMRs. Here, in order to better compare the capture mechanism with the limiting cases and give more accurate results, we choose to use the non-integrable Hamiltonian. A thorough study of the phase space topology is difficult; nevertheless, we can gain
considerable insight into the resonant motion of the two planets by
analyzing the fixed points of the Hamiltonian.
}

For two planets with comparable masses in a second-order MMR,
the Hamiltonian can be simplified to that of a two degree of freedom
system (see Appendix A)
\eq{\label{Ham_simMass}
-H =& (x_1^2+x_2^2+y_1^2+y_2^2+\eta)^2 + (Ax_1+Bx_2)^2 \\
&+ (Cy_1+Dy_2)^2 +E^2x_1^2,
}
where the canonical momentum and coordinate pairs $(x_1,y_1)$ and
$(x_2,y_2)$ are defined by (for $e_i\ll 1$)
\eal{
\label{xyDef_1}&x_1 = \sqrt{2\Theta_1}\cos \theta_1,~~~y_1 = \sqrt{2\Theta_1}\sin \theta_1,\\
\label{xyDef_4}&x_2 = \sqrt{2\Theta_2}\cos \theta_2,~~~y_2 = \sqrt{2\Theta_2}\sin \theta_2,
}
with
\eq{
\Theta_1=\frac 1 2\mu_2^{-1}\alpha_{\rm res}^{1/2}e_1^2,~~~\Theta_2=\frac 1 2\mu_1^{-1}e_2^2,
}
where $\alpha_{\rm res}\equiv [(j-2)/j]^{2/3}$. The resonant angles $\theta_1,\theta_2$ are
\eal{
&\theta_1 = \frac j 2 \lambda_2 - \frac{(j-2)}{2}\lambda_1 -\varpi_1,\\
&\theta_2 = \frac j 2 \lambda_2 - \frac{(j-2)}{2}\lambda_1 -\varpi_2.
}
The parameter $\eta$ is a constant in the absence of dissipation, and is related to $\alpha=a_1/a_2, e_1$ and $e_2$ by
\eq{\label{eta_simMass}
\eta = &\frac{4}{j\mu_0}\left[1-\left(\frac{\alpha}{\alpha_{\rm res}}\right)^{1/2}\right]-\mu_2^{-1}\alpha_{\rm res}^{1/2}e_1^2\\
&-\mu_1^{-1}e_2^2+\text{constant of $\mathcal O(1)$},
}
where
\eq{
\mu_0\equiv \mu_1+\mu_2\alpha_{\rm res}
\label{eq:muodefine}}
is an ``effective total mass ratio" that will frequently appear in
this section, and the $\mathcal O(1)$ constant can be solved numerically (see Appendix A). Note that the last term in Eq.~\eqref{eta_simMass} trivially
shifts the location of the resonance, with the shift in resonant
$\alpha$ being of order $\mu_0$. For the Hamiltonian
\eqref{Ham_simMass}, time is normalized by
\eq{
\tau=t/T_0,~~~T_0\equiv \left(\frac {3j^2} {32}\mu_0n_2\right)^{-1}\sim \mu_0^{-1}P_1.
}
The parameters $A,B,C,D,E$ are all real, and they only depend on $j$ and the mass ratio
\eq{
q\equiv m_1/m_2.
}
For $q\sim 1$, these parameters are all of order unity. The full derivation of the Hamiltonian can be found in Appendix A.

\begin{figure}
\centering
\includegraphics[width=\columnwidth]{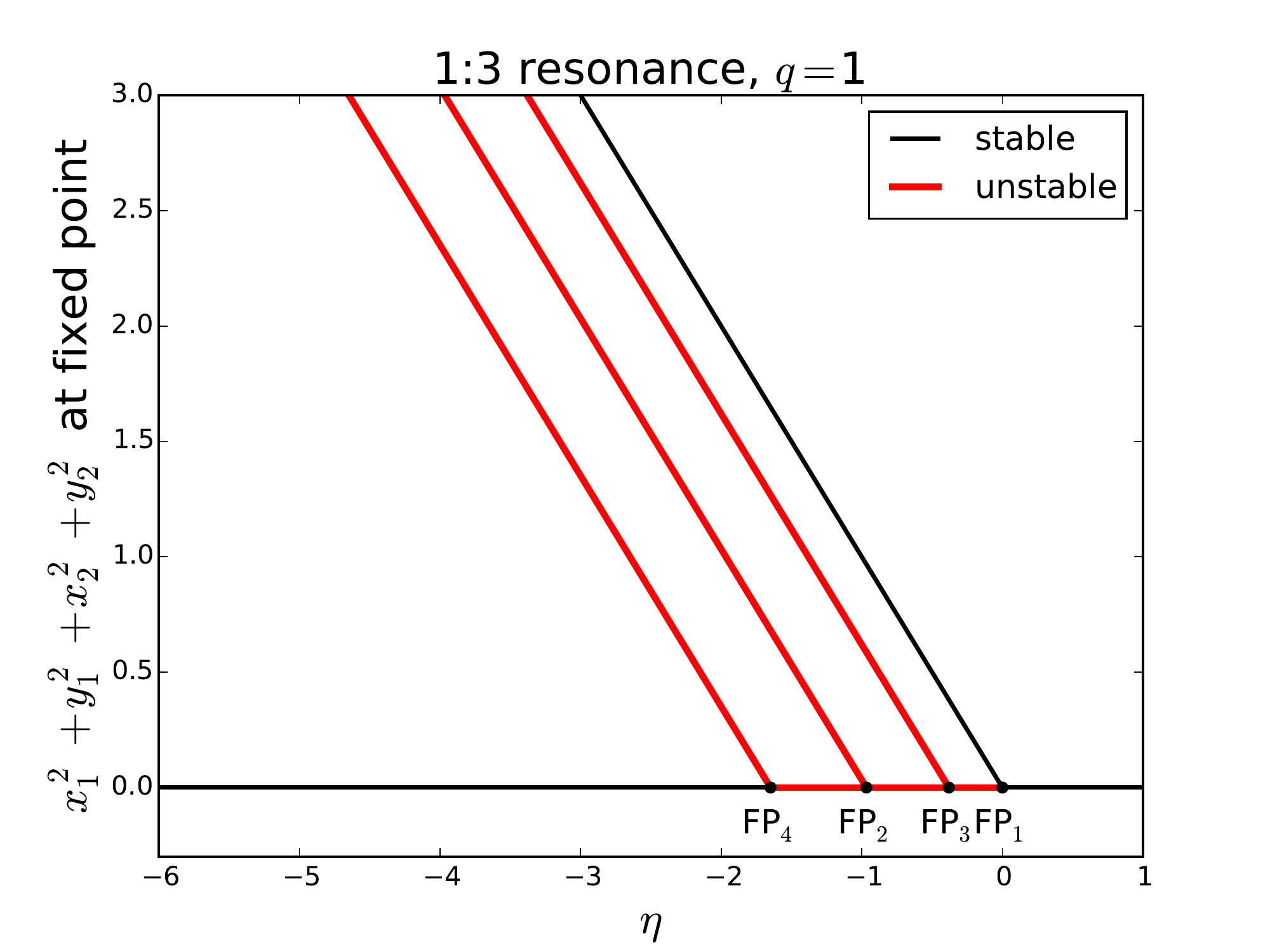}\\
\includegraphics[width=\columnwidth]{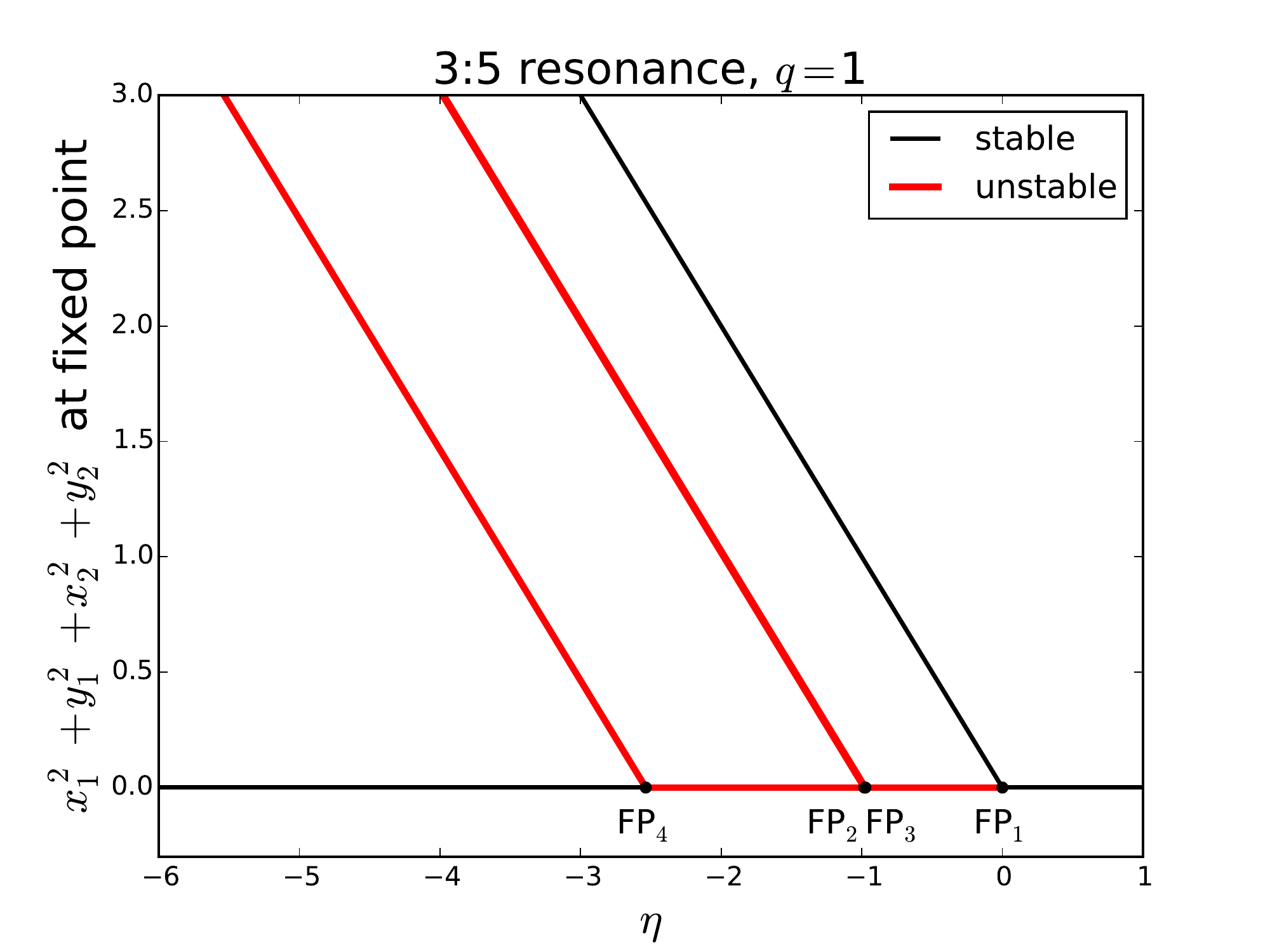}
\caption{Locations of fixed points for 1:3 and 3:5 resonance with
  equal mass planets ($m_1=m_2$). The vertical axis shows the
  (squared) distance between the fixed point and the origin (Eq.~\ref{eq:dist2}).
 The four pairs of fixed points other than the origin (FP$_0$) are labeled by
  ${\rm FP}_1$ - ${\rm FP}_4$. Note that each branch of FP$_i$
  corresponds to two fixed points; and all fixed points move away from
  the origin along different directions as $\eta$ decreases. The
  stability of the fixed points are marked by color; red intervals are
  unstable. We see that the stability for each branch of fixed points
  and the distribution of branching points are qualitatively the same,
  except that for the 3:5 resonance the branching points of FP$_2$ and
  FP$_3$ are much closer. Results for resonances with larger $j$
  (e.g. 5:7, 7:9) and other mass ratios are qualitatively the
  same. Note that the distribution of the stable fixed points is very
  similar to that in Figure \ref{fixedPtsLoc_toy}.}
\label{fixedPtsLoc}
\end{figure}

The phase space now have 4 dimensions, so we cannot conveniently plot
the phase space topology as we did for the restricted
problem in Sections 2-3.  Instead, we analyze the fixed points of the system, which are
informative enough to illustrate the resonance capture mechanism. The
coordinates of the fixed points can be calculated by solving
$d(x_1,x_2,y_1,y_2)/d\tau=0$ (see Appendix A). The system can have
$1,3,5,7$ or $9$ fixed points. There is always a fixed point that lies
in the origin (labeled as FP$_0$), and all other fixed points come in
pairs [i.e. if $(x_{10},x_{20},y_{10},y_{20})$ is a fixed point, then
  $(-x_{10},-x_{20},-y_{10},-y_{20})$ is also a fixed point, and their
  properties are identical]. These fixed points have the form:
\eal{\label{FP1}
&{\rm FP}_1: ~~~(0,0,\pm y_{11}\sqrt{-\eta},\pm y_{21}\sqrt{-\eta})\\
&{\rm FP}_2: ~~~(0,0,\pm y_{12}\sqrt{-\eta+\eta_2},\mp y_{22}\sqrt{-\eta+\eta_2})\\
&{\rm FP}_3: ~~~(\pm x_{11}\sqrt{-\eta+\eta_3},\pm x_{21}\sqrt{-\eta+\eta_3},0,0)\\
&{\rm FP}_4: ~~~(\pm x_{12}\sqrt{-\eta+\eta_4},\pm x_{22}\sqrt{-\eta+\eta_4},0,0)
}
where the parameters $\eta_{2,3,4}$ and $x_{ij}$, $y_{ij}$ are functions of $A,B,C,D,E$, and\footnote{The relation between $\eta_{2,3,4}$ in \eqref{branching} come from a numerical survey across the parameter space, while we can analytically prove $\eta_{2,3,4}<0$.}
\eq{\label{branching}
\eta_4<\eta_2<\eta_3<0,
}
\eq{
y_{11}^2+y_{21}^2=y_{12}^2+y_{22}^2=x_{11}^2+x_{21}^2=x_{12}^2+x_{22}^2=1.
}
FP$_1$ exists only for $\eta<0$, and FP$_i$ ($i=2,3,4$) only for
$\eta<\eta_i$. Note that $\eta=\eta_4$ marks one boundary of the
resonant region, while the other boundary is $\eta=0$. The $\eta$
value at which different fixed points ``branch out" from the origin,
and the distances of the fixed points from the origin, are summarized
in Figure \ref{fixedPtsLoc}.
Note that the (squared) distance is related to the planet's eccentricities by 
\eq{
x_1^2+y_1^2+x_2^2+y_2^2=\mu_2^{-1}\alpha_{\rm res}^{1/2}e_1^2+\mu_1^{-1}e_2^2.
\label{eq:dist2}}

The stability of fixed points FP$_1$ is manifest, since they are local
maxima of the Hamiltonian. Also, FP$_0$ is stable when the system is
away from resonance ($\eta<\eta_4$ or $\eta>0$),
since it is clearly an extremum of the Hamiltonian when $|\eta|\gg 1$.
In general, the stability of a fixed point can be determined by
calculating the eigenvalues of the linearized equations of motion
(similar to the stability analysis for Section 2.4 and 3.3, except
that the system is non-dissipative); a fixed point is stable only when
all eigenvalues have zero real part\footnote{Note that a stable fixed
  point of a non-dissipative system cannot have any negative
  eigenvalue because when all eigenvalues are non-positive and some
  are negative, the phase space volume of the flow is not
  conserved.}. We numerically computed the eigenvalues for each fixed
point, and the result is shown in Figure \ref{fixedPtsLoc}. It turns out that
FP$_2$, FP$_3$ and FP$_4$
are unstable; and FP$_0$ is unstable when $\eta\in(\eta_4,0)$. The
fixed point FP$_1$ is always stable.

Comparing Figure \ref{fixedPtsLoc} with Figure \ref{fixedPtsLoc_toy}
shows that the locations of the stable fixed points for $q\sim 1$ are
qualitatively the same as in the limiting mass ratio cases\footnote{\label{eta_footnote}In the limit of $q\ll1$ or $q\gg1$, the definition of $\eta$ in Eq.\eqref{eta_simMass} differs from that used in Sections 2 and 3 by unity. Thus in Fig.8, the system enters resonance at $\eta=0$ as $\eta$ decreases, while in Fig.3 this occurs at $\eta=1$.}, while
extra branches of unstable fixed points exist for $q\sim
1$. Therefore, the mechanism of resonance capture and eccentricity
excitation should be mostly similar: 

(i) For slowly increasing $\eta$, 
the eccentricities of the planets suddenly increase as the origin becomes unstable
at $\eta=\eta_4$; after that the eccentricities remain approximately constant, 
conserving the phase space volume. The eccentricities can be excited to 
$\Theta_1\sim \Theta_2\sim 1$,
and the system is not captured into resonance (both resonant angles are circulating). 

(ii) For slowly decreasing $\eta$, the system tends to follow the stable fixed point,
so it begins to advect with FP$_1$ as $\eta$ passes 0 and enters the
resonant zone. The eccentricities keep growing with decreasing $\eta$;
the system may reach an equilibrium if $\eta$ becomes constant due to
the dissipative effect associated with planet-disk interaction.

\subsection{Capture mechanism and condition}

To examine the effect of planet migration we need to consider
$d\eta/dt$. Using Eqs.~\eqref{diss_1}, \eqref{diss_2} and
\eqref{eta_simMass}, we have
\eq{\label{detadtau}
\frac{d\eta}{dt} = &-\frac{2}{j\mu_0}\left(T_{m,2}^{-1}-T_{m,1}^{-1}\right)\\
&+2T_{e,1}^{-1}\Theta_1\left[1+\frac{p_1}{(j-2)(1+\alpha_{\rm res}^{-1}q)}\right]\\
&+2T_{e,2}^{-1}\Theta_2\left[1-\frac{p_2}{j(1+\alpha_{\rm res}q^{-1})}\right].
}
For the following calculation, we will assume $p_1=p_2=2$. The equations of motion for $x_i,y_i$ can be obtained by adding dissipative terms to the non-dissipative equations [derived from the Hamiltonian \eqref{Ham_simMass}] for $dx_i/d\tau$ and $dy_i/d\tau$. These dissipative terms are
\eq{
\left.\frac{dx_i}{d\tau}\right|_{\rm diss} = -\frac{x_i}{\tau_{e,i}},~~~\left.\frac{dy_i}{d\tau}\right|_{\rm diss} = -\frac{y_i}{\tau_{e,i}}.
}

\begin{figure*}
\centering
\includegraphics[width=.8\textwidth]{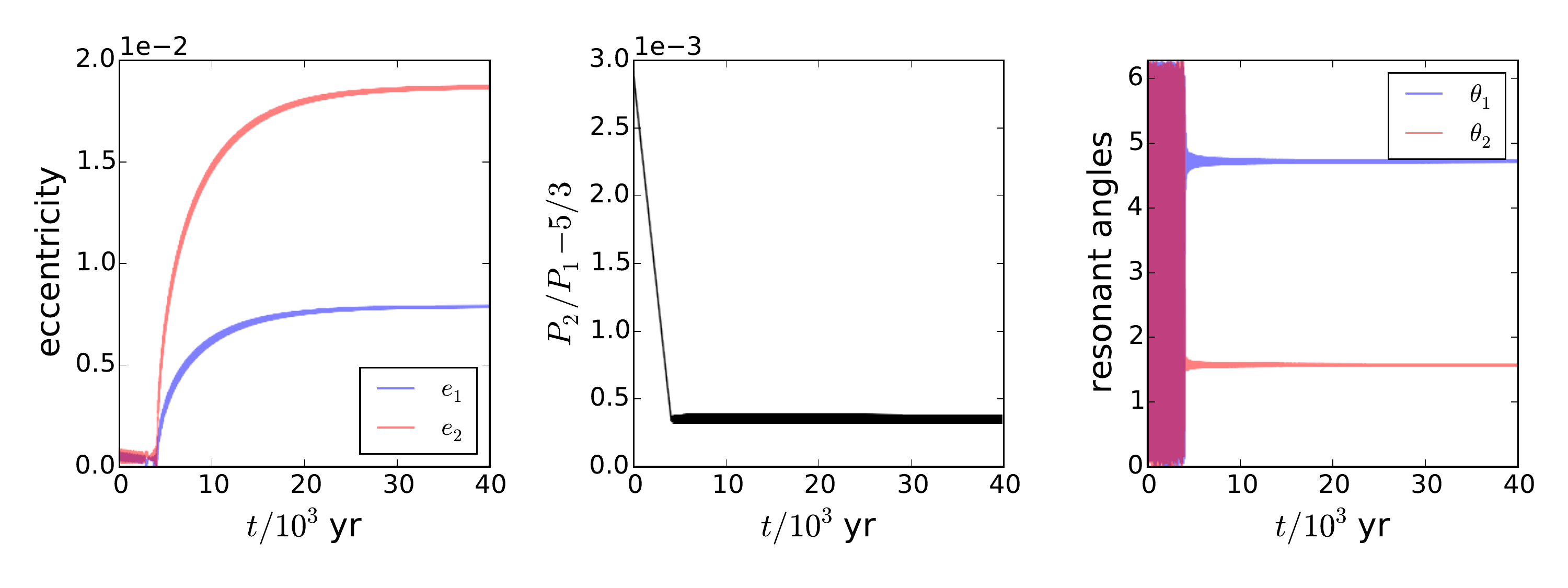}
\caption{Stable capture of similar mass planets into 3:5 MMR. The
  system has $M_\star=1M_\odot, ~~m_1=10M_\oplus, ~~m_2=5M_\oplus,
  ~~a_1=0.1AU, ~~T_{m,1}=4\,{\rm Myr}, ~~T_{m,2}=2\,{\rm Myr}$, and
  $T_e=10\,{\rm kyr}$. We see that the eccentricities of both planets
  are excited, and the period ratio, eccentricities and resonant
  angles all converge to constants. The two resonant angles are at
  $\theta_1=\pi/2$ and $\theta_2=3\pi/2$. Note that another final
  state with $\theta_1$ and $\theta_2$ switched is equally likely to
  occur.}
\label{simMassStb}
\end{figure*}

\begin{figure*}
\centering
\includegraphics[width=.8\textwidth]{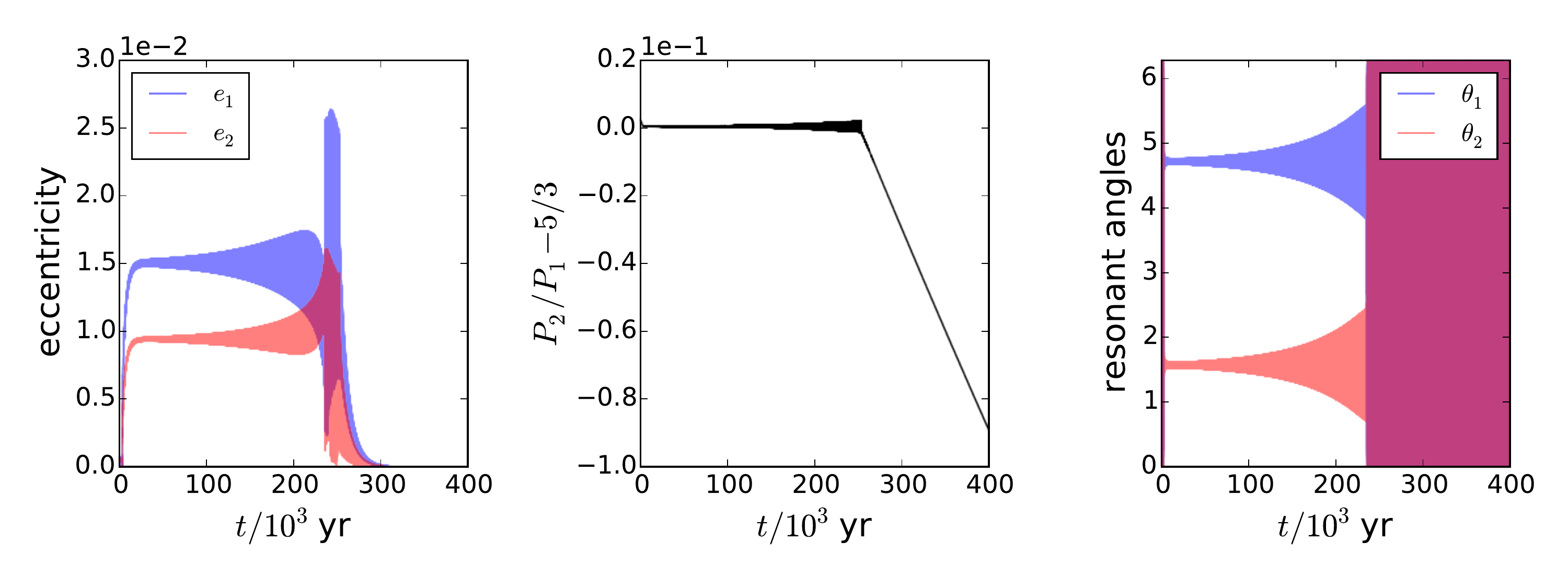}
\caption{Capture of similar mass planets into 3:5 MMR followed by
  escape from the resonance (overstable equilibrium with $\eta_{\rm eq}<\eta_4$).
 The system is the same as that in Fig.~\ref{simMassStb}, except
 $m_1=5M_\oplus$ and $m_2=10M_\oplus$.  We see that after reaching the
 equilibrium state, the libration amplitudes of resonant angles
 increase due to overstability and the system escapes from resonance
 eventually.}
\label{simMassOverstb1}
\end{figure*}

\begin{figure*}
\centering
\includegraphics[width=.8\textwidth]{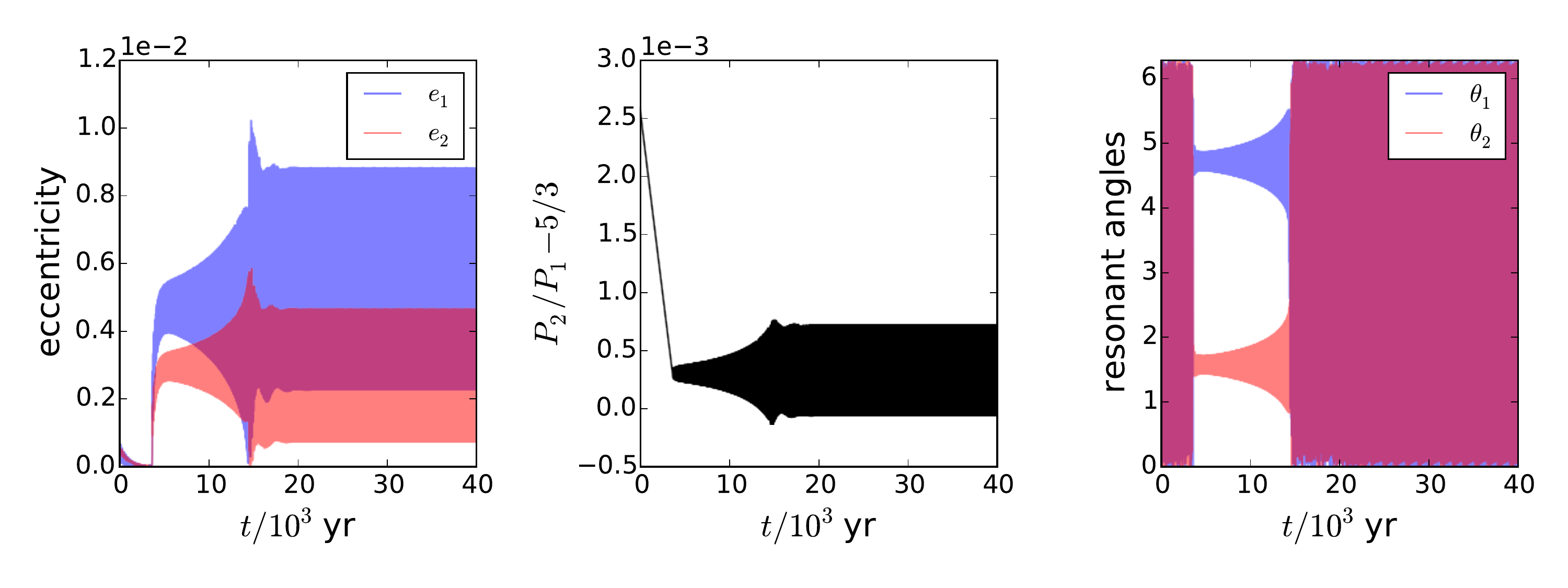}
\caption{Capture of similar mass planets into 3:5 MMR followed by
  ``partial'' escape (overstable equilibrium with
  $\eta_{\rm eq}>\eta_4$).
The system is the same as that in Fig.~\ref{simMassOverstb1}, except
$T_e=1~{\rm kyr}$. The libration around equilibrium is still
overstable, but the system ends up in a stable circulating orbit with
the period ratio oscillating around a constant value.}
\label{simMassOverstb2}
\end{figure*}

Resonant capture requires $d\eta/d\tau<0$ (see the last paragraph of
Section 5.1). Thus a necessary condition for capture is
\eq{
T_m\equiv \frac{1}{T_{m,2}^{-1}-T_{m,1}^{-1}}>0.
}
In other words, resonance capture always requires convergent migration,
in agreement with the limiting mass cases. When the eccentricities of
the planets are excited due to the decrease of $\eta$, the increased
eccentricities tend to increase $\eta$ because the last two terms in
Eq.~\eqref{detadtau} are always positive. Therefore, there exists an
equilibrium state where the decrease of $\eta$ due to convergent
migration is balanced by the increase of $\eta$ due to excited
eccentricities. The system can be held at this equilibrium state
permanently if small oscillations about the equilibrium point is
stable (see Fig.~\ref{simMassStb}). When the equilibrium point is
overstable, the system will eventually leave the libration zone. In
this case, the system exhibits two possible outcomes, depending on
$\eta_{\rm eq}$ (the value of $\eta$ at the equilibrium state): For
$\eta_{\rm eq}<\eta_4$ ($\sim -1$; see Fig.~\ref{fixedPtsLoc}),
the system escapes from the
resonance, and the period ratio increasingly deviates from the
resonant value (Fig. \ref{simMassOverstb1}); 
for $\eta_{\rm eq} > \eta_4$, the system eventually enters a
quasi-equilibrium state, where the resonant angles circulate but the
period ratio stays approximately constant and the eccentricities
undergo oscillations (see Fig.~\ref{simMassOverstb2}).
The stability criterion of the equilibrium state following resonance capture is discussed
in Section 5.4.

Comparing these results with the two limiting cases ($q\ll1, q\gg 1$)
shows that the mechanism of resonance capture is qualitatively the
same: The systems starts with small eccentricities; as $\eta$
decreases and passes 0, the system advects with (one of) FP$_1$
because the origin is no longer stable; further evolution of the
system after capture depends on the ``location" ($\eta_{\rm eq}$) and
stability of the equilibrium state. We expect that the requirement on
$T_e$ and $T_m$ for resonance capture should be similar to the
limiting cases: For $T_e$, we require
\eq{
T_{e,i}\gtrsim T_0.
}
Note that the system can still be captured into resonance if one of
$T_{e,i}$ ($i=1,2$) is too small to allow excitation of $e_i$; in this
case only the eccentricity of the other planet is excited and only
that planet's resonant angle is librating upon capture.
Similar to Eq.~(\ref{trap_cond_1}), the requirement on $T_m$ is
\eq{
T_m\gtrsim \mu_0^{-1}T_0|\ln\Theta_{i,0}|,
}
where $\Theta_{i,0}$ is the initial (prior to entering the resonance)
value of $\Theta_i$.

\subsection{Effect of large initial eccentricities}

So far in this section we have assumed that the initial eccentricities of the two
planets are small ($\Theta_{1,0} \sim e_{1,0}^2/\mu_2\lesssim 1$ and
$\Theta_{2,0}\sim e_{2,0}^2/\mu_1\lesssim 1$). This assumption is in
general satisfied for small planets because their eccentricities can be
quickly damped by the disk. 
In systems consisting of a massive planet and a very small planet, the dimensionless
eccentricity parameter ($\Theta$) of 
the big planet is inversely proportional to the smaller planet's mass 
and can become large even when the physical eccentricity is small. Here we
consider the effect when one or both planets have $\Theta_{i,0}\gtrsim 1$.

As noted in Section 4, for the restricted three-body problem, resonance trapping
becomes probabilistic when $\Theta_0\gtrsim 1$ even for very large 
(infinite) $T_e$ and $T_m$.
For systems with comparable mass planets, we can similarly conjecture
that trapping becomes probabilistic if $\Theta_{1,0}$ or
$\Theta_{2,0}$ is $\gtrsim 1$. This is because when the initial
$\Theta_1$ or $\Theta_2$ is $\gtrsim 1$, the system cannot settle near
the stable fixed point FP$_1$ before the origin (FP$_0$) becomes
stable again, and resonance trapping should be
probabilistic. Numerical integration of the equations of motion when
one or both planets have initial $\Theta_i\gtrsim 1$ confirms this conjecture.

This line of reasoning does not fully apply to systems with relatively
large mass ratio ($q$ or $q^{-1}\gtrsim 5$). For such systems, the
perturbation on the larger planet from the smaller one is much weaker
than the perturbation on the smaller planet from the larger one;
therefore the evolution of the small planet's eccentricity is barely
affected by the other planet's eccentricity.
Thus, if the smaller planet has a large initial $\Theta$, it cannot be captured into resonance.
However, when only the larger planet has initial $\Theta \gtrsim 1$,
the small planet can still be captured into resonance; in this case
the small planet attains an excited eccentricity and its corresponding
resonant angle librates, while the larger planet's eccentricity is
not excited, and its resonant angle circulates (see Sections 2-4).

\subsection{Stability of capture and escape from resonance}

As discussed in Section 5.2, after resonance capture, the system
approaches an equilibrium in which eccentricity excitation due to
resonant forcing balances eccentricity damping due to planet-disk
interaction. Further evolution of the system depends on the location
of the equilibrium and its stability. In particular, if the
equilibrium is overstable, the system will completely escape from the
resonance if 
$\eta_{\rm eq} < \eta_4\sim -1$, 
where $\eta_{\rm eq}$ is the value of $\eta$ in the equilibrium state. The
equilibrium state can be determined by setting
$dx_i/d\tau=dy_i/d\tau=0$ (including the dissipative terms) and
$d\eta/d\tau=0$ [see Eq. \eqref{detadtau}]. In orders of magnitude, we find\footnote{Note that this value of $\eta$ is shifted from Eq. \eqref{eq:etaeq1} or Eq. \eqref{eta_def} by unity. See Footnote \ref{eta_footnote}.}
\eq{\label{eta_eq_scale}
|\eta_{\rm eq}| \sim \frac{T_{m,2}^{-1}-T_{m,1}^{-1}}{\mu_0(T_{e,1}^{-1}+T_{e,2}^{-1})},}
and when $q\sim 1$,
\eq{
\Theta_{1,\rm eq}\sim \Theta_{2,\rm eq}\sim |\eta_{\rm eq}|.
}
The resonant angles are $\theta_1\simeq \pm \pi/2$ and $\theta_2\simeq \mp\pi/2$.

Based on the results of the restricted three-body problems (Sections 2-3), we expect resonance
capture (equilibrium) to be stable when $q\equiv m_1/m_2\gtrsim 1$, and
overstable when $q\lesssim 1$. 
Here we provide a more accurate determination of the critical
$q=q_{\rm crit}$ at which the transition from stability to overstability occurs,
in order to predict the overall population of planet pairs in
second-order resonances as a result of convergent migration. 
Note that we only examine capturing into the equilibrium state where both
resonant angles librate. Capture into resonance with one
librating resonant angle (e.g. the $q\to 0$ or $\infty$ limiting
cases) or capture with no librating resonant angle (i.e. when the
equilibrium point is unstable but $\eta_{\rm eq}\gtrsim \eta_4$) are
not considered.

To determine the stability of resonance capture, we linearize the
equations of motion (with dissipation) near the equilibrium point and
calculate the eigenvalue $\lambda$. For an overstable system,
the escape timescale from the resonance is
\eq{
\tau_{\rm esc}\sim 1/\lambda_r~~~(\text{when }\lambda_r>0)
}
where $\lambda_r$ is the maximum real part of the eigenvalues.

Because of the complexity of the system, we cannot obtain a simple
analytical expression for $\lambda_r$ as we did for the restricted three-body
problems (Sections 2-3); instead, we solve the eigenvalue problem numerically to
calculate $\tau_{\rm esc}$ for various parameters. 
The system is fully specified by the parameters
$\tau_{e,i},\tau_m,\mu_i,n_1,j$ and $q$. From the dimensionless form
of the equations of motion we see that all factors of $n_1$ cancel
out, and $\mu_i$ and $\tau_m$ only appear in the combination
$\mu_0\tau_m$ (see Appendix A). For clarity, we use
\eq{
\tau_e \equiv \sqrt{\tau_{e,1}\tau_{e,2}},~~~q_e = \tau_{e,2}/\tau_{e,1}
}
instead of $\tau_{e,i}$. Note that $q_e$ is the ratio of the eccentricity damping rates. In
general, we find that $\tau_{\rm esc}$ depends very weakly on
$\mu_0\tau_m$, as long as $\mu_0\tau_m\lesssim \tau_e$, in agreement
with the results of the restricted problem.\footnote{Note that
  realistic systems usually have $\mu_0\tau_m\lesssim \tau_e$. When
  $\mu_0\tau_m\gtrsim \tau_e$, which is less common but still
  possible, the period ratio can be permanently hold near resonance
  regardless of the stability of the equilibrium point, since in this
  regime $|\eta_{\rm eq}|\lesssim 1$.} Thus, we only need to consider
the dependence of $\tau_{\rm esc}$ on $\tau_{e},q_e,j$ and $q$.

\begin{figure}
\centering
\includegraphics[width=\columnwidth]{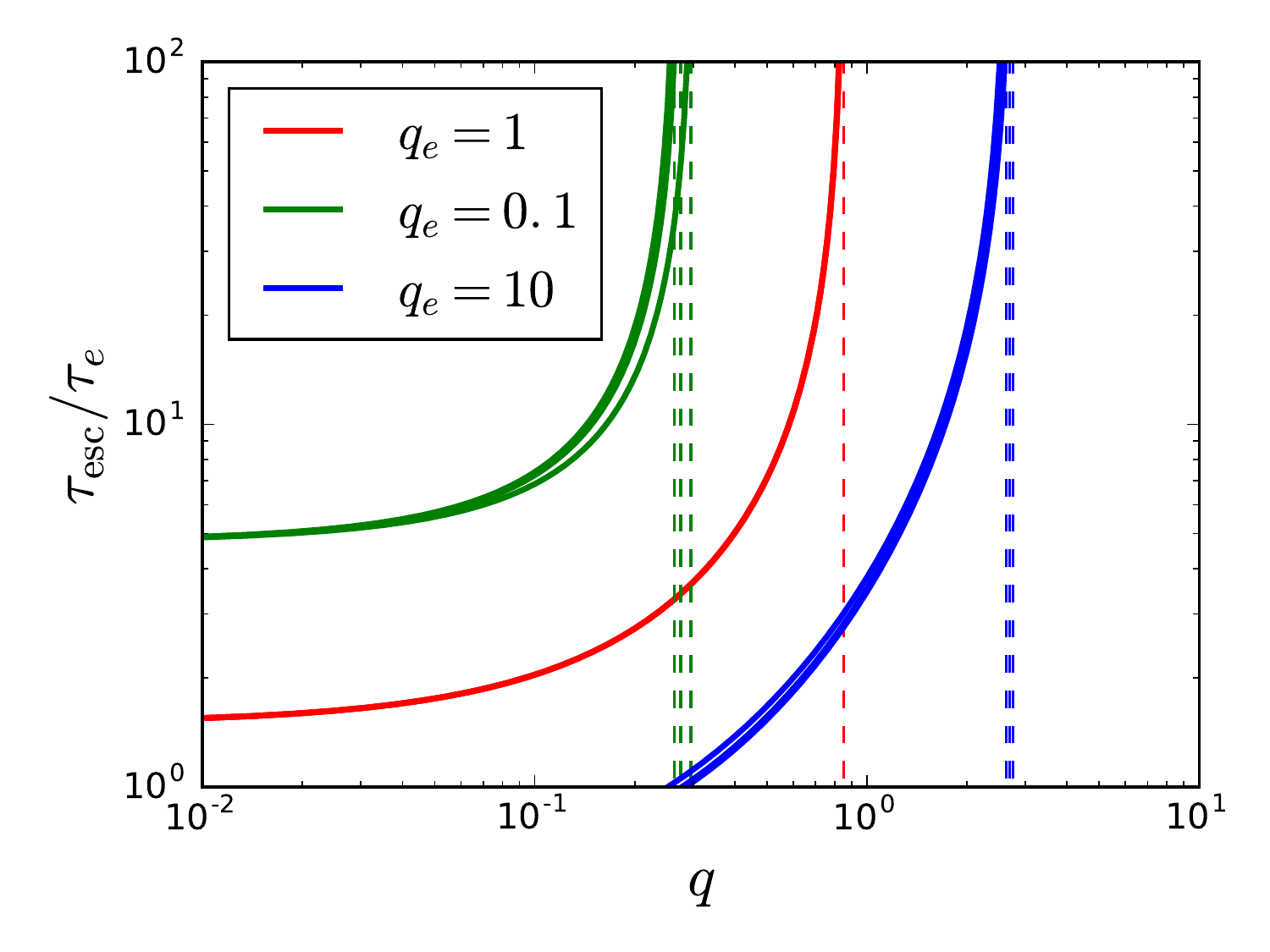}
\caption{Resonance escape time $\tau_{\rm esc}$ (in units of $\tau_e$,
  the eccentricity damping time) for the 3:5 MMR as a function of
  $q=m_1/m_2$ for different $\tau_e$ and $q_e=\tau_{e,2}/\tau_{e,1}$,
  with given $\mu_0\tau_m=10$. Different colors mark different $q_e$;
  for each $q_e$ there are three curves for $\tau_e=100,300$ and
  $1000$. The curves are calculated using $\mu_0\tau_m=100$, although
  the results are unchanged for other values of $\mu_0\tau_m\lesssim
  \tau_e$. We see that for each $q_e$, the difference between the
  three curves are very small, suggesting that $\tau_{\rm esc}/\tau_e$
  barely depends on $\tau_e$. In general $\tau_{\rm esc}/\tau_e$ is
  smaller for larger $q_e$, but $q_{\rm crit}$ is larger. Other
  resonances show similar trends.}
\label{stability_te}
\end{figure}

\begin{figure}
\centering
\includegraphics[width=\columnwidth]{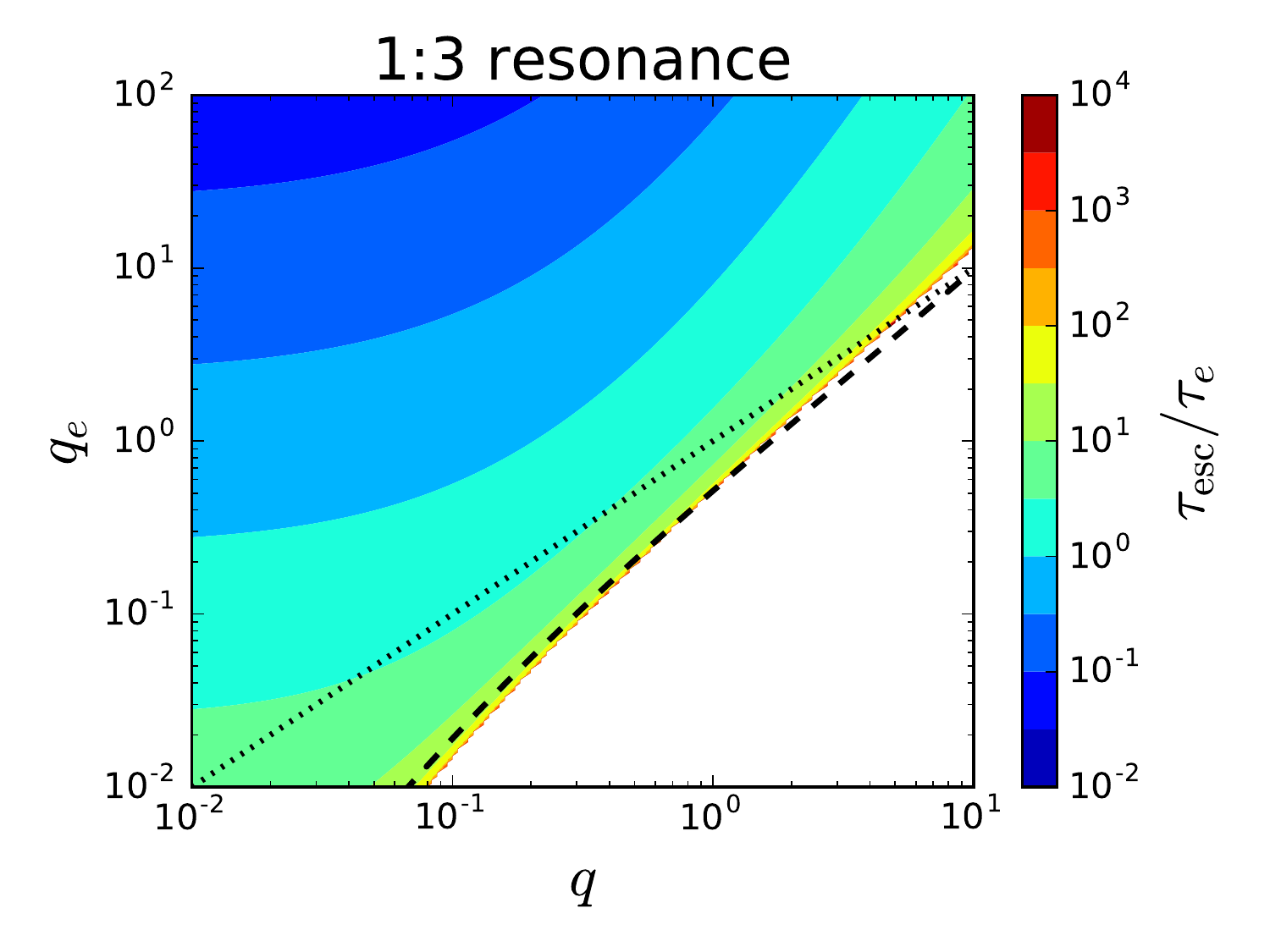}
\includegraphics[width=\columnwidth]{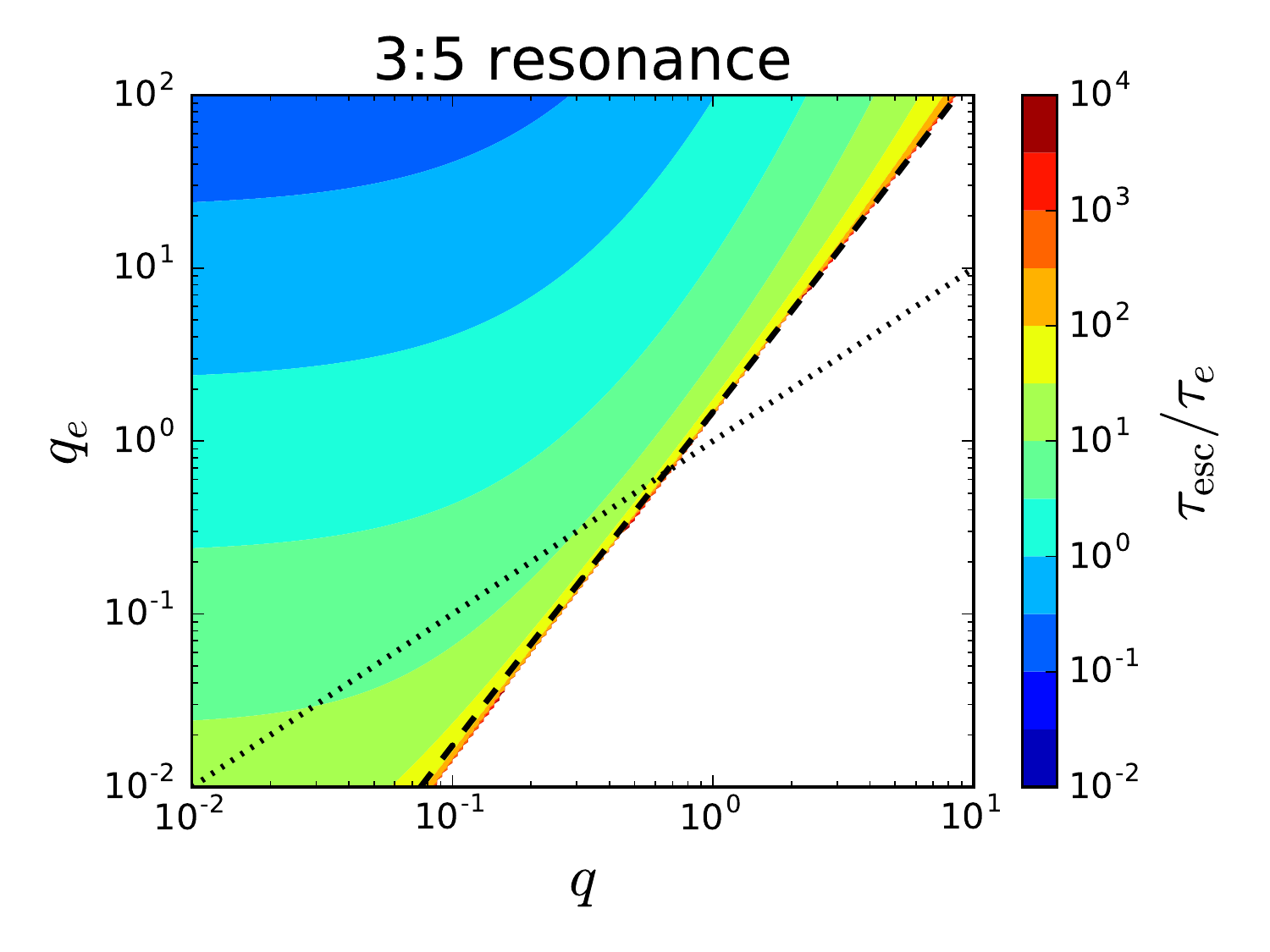}
\includegraphics[width=\columnwidth]{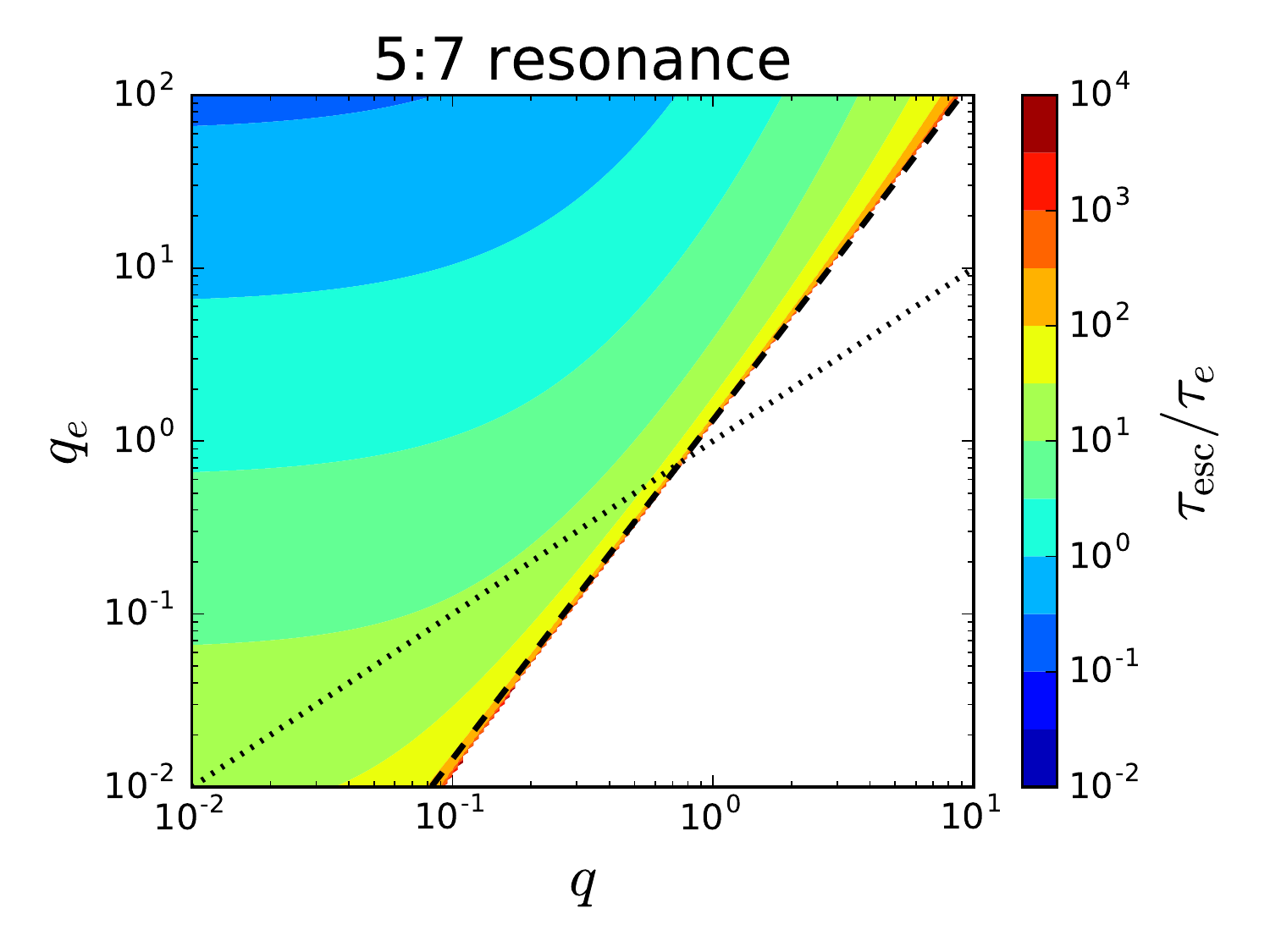}
\caption{Resonance escape time $\tau_{\rm esc}$ for the 1:3, 3:5 and
  5:7 MMRs as a function of $q=m_1/m_2$ and
  $q_e=\tau_{e,2}/\tau_{e,1}$. The results are obtained using the
  values $\mu_0\tau_m=10$ and $\tau_e=1000$, although the results are
  nearly unchanged for other values as long as $\mu_0\tau_m\lesssim
  \tau_e$. Color shows the value of $\tau_{\rm esc}/\tau_e$, while the
  white region corresponds to stability (infinite escape time). Note that the behavior of the
  3:5 and 5:7 resonances are very similar, while for the 1:3 resonance
  the overstable region is larger. 
The stability limit is approximately $q_{\rm crit}\simeq q_e^{1/2}$ for the 3:5 and 5:7
  resonances, and 
$q_{\rm crit}\simeq 2 q_e^{2/3}$ for the 1:3
  resonance. 
  {The black dashed line shows the stability boundary using the criterion of \citet{Delisle15} (and using the relationship between the eccentricity ratio and mass ratio as depicted in Figure \ref{eccRatio}), which agrees with our result.} 
  The black dotted line shows $q_e=q$; systems with
  low-mass planets in a gaseous disk should lie close to this line. For
  such systems, $q_{\rm crit}$ for the 3:5 or 5:7 resonance is smaller
  than that for the 1:3 resonance.}
\label{stability_qe}
\end{figure}

Figure \ref{stability_te} 
shows $\tau_{\rm esc}=\lambda_r^{-1}$ for the 3:5 MMR as a function of
$q$ for different values of $q_e$ and $\tau_e$. We see that $\tau_{\rm esc}/\tau_e$ 
depends weakly on $\tau_e$ (similar to the behavior of
the restricted problem). In general, $\tau_{\rm esc}/\tau_e$ decreases
with increasing $q_e$, while the critical $q_{\rm crit}$ increases
with $q_e$. Thus, for a given MMR (given $j$), the stability of the
equilibrium mainly depends on $q$ and $q_e$. Figure \ref{stability_qe}
shows this dependence for different resonances. Empirically, we find
that the critical mass ratio (above which the equilibrium state is stable)
is given by
\eal{
\label{qcrit3}&q_{\rm crit}\simeq 2q_e^{2/3} &&\text{for 1:3 resonance,}\\
\label{qcrit5}&q_{\rm cirt}\simeq q_e^{1/2} &&\text{for other resonances.}
}
The result for the 1:3 MMR is different because the corresponding
$\alpha_{\rm res}$ is much smaller than that of the other resonances,
and there is an ``indirect term" (see \citealt{MurrayDermott99}) that
only appears in the 1:3 MMR. We see that for most of the unstable
region in the $q-q_e$ parameter space, $\tau_{\rm esc}$ is close to
$\tau_e$. Near the stability limit, $\tau_{\rm esc}$ is significantly
larger than $\tau_e$. For $q\lesssim 1$, we find that $\tau_{\rm esc}\sim\tau_{e,1}$.

\begin{figure}
\centering
\includegraphics[width=\columnwidth]{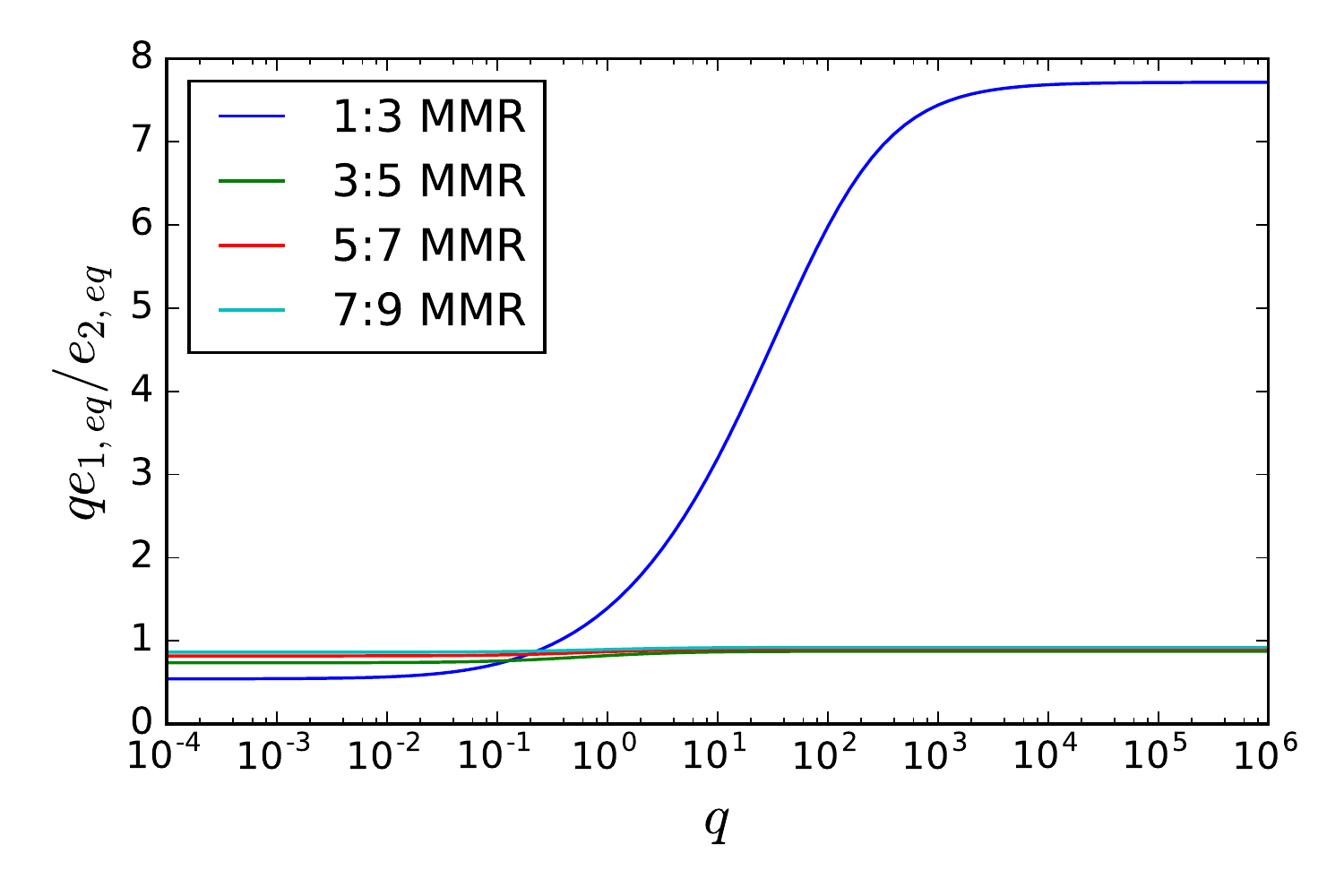}
\caption{{Ratio of the two planets' eccentricities at the equilibrium state (the fixed point FP$_1$) as a function of the mass ratio $q$. For $q\ll1$ and $q\gg 1$, $qe_{1,\rm eq}/e_{2,\rm eq}$ is nearly independent of $q$. For $q\sim 1$, $qe_{1,\rm eq}/e_{2,\rm eq}$ grows significantly as $q$ increases for the 1:3 MMR and is nearly constant for the other second-order MMRs. This explains the difference between equations \eqref{qcrit3} and \eqref{qcrit5}.}}
\label{eccRatio}
\end{figure}

{
\citet{Delisle15} gives the stability critarion in terms of the equilibrium eccentricity ratio; the captured state is stable if $q_e^{-1}\gtrsim (e_{1,\rm eq}/e_{2,\rm eq})^2$. Figure \ref{eccRatio} shows the eccentricity ratio at the captured equilibrium state, which we have analytically calculated using the location of the fixed point FP$_1$. We see that for all second-order MMRs except the 1:3 MMR, $e_{1,\rm eq}/e_{2,\rm eq}\simeq q^{-1}$; for 1:3 MMR $e_{1,\rm eq}/e_{2,\rm eq}$ has a different dependence on $q$ around $q\sim 1$. This explains why our stability boundary \eqref{qcrit3} is different only for the 1:3 MMR. For second-order MMRs other than the 1:3 MMR, the criterion in \citet{Delisle15} suggests that capture is stable for $q_e\lesssim q^2$, agreeing with our equation \eqref{qcrit5}.
Figure \ref{stability_qe} compares $q_{\rm crit}$ obtained using the criterion of \citet{Delisle15} with our result; we see that the two results are very similar when the relationship between the eccentricity ratio and mass ratio (Figure \ref{eccRatio}) is spelled out.
}

For realistic systems with relatively small (non-gap-opening) planets, we have [see Eq. \eqref{Te}]
\eq{\label{qeq}
\frac{q_e}{q}\sim\frac{F(a_1)}{F(a_2)}~~\text{with}~~F(a)\equiv \Sigma_da^{4.5}h^{-4},
}
where $\Sigma_d$ and $h$ are surface density and scale height of the disk. Since the two planets in resonance are relatively close to each other, we expect this ratio to be close to 1. Therefore, the system should lie near the black dashed line in Figure \ref{stability_qe}. From this figure we see that for systems with $q_e=q$, $q_{\rm crit}$ is slightly below 1 for the 3:5 and 5:7 resonances, and about 5 for the 1:3 resonance. This means that a system undergoing convergent migration (which usually implies $q,q_e\lesssim 1$) is unlikely to be stably captured into the 1:3 resonance, while stable capture into the 3:5 or 5:7 resonance is possible.

Note that $q_{\rm crit}$ is sensitive to the physical property of the disk,
especially for the 1:3 resonance. For disks with $F(a_1)>F(a_2)$, we
expect $q_e/q>1$, and $q_{\rm crit}$ is shifted to larger values
(compared to that depicted on Figure \ref{stability_qe}); for disks
with $F(a_1)<F(a_2)$, $q_{\rm crit}$ is shifted to smaller
values. Using Eqs.~\eqref{qeq} and \eqref{qcrit3}-\eqref{qcrit5}, we
find $q_{\rm crit}\simeq 8[F(a_1)/F(a_2)]^2$ for the 1:3 resonance and
$q_{\rm cirt}\simeq F(a_1)/F(a_2)$ for the other resonances. Also
recall that convergent migration requires $T_{m,1}^{-1}<T_{m,2}^{-1}$,
which implies $q\lesssim
(\Sigma_da^{2.5}h^{-2})_2/(\Sigma_da^{2.5}h^{-2})_1$ using
Eq.~\eqref{Tm}. This again suggests that the stable capture into the
1:3 resonance during convergent migration is unlikely.

\begin{figure}
\centering
\includegraphics[width=\columnwidth]{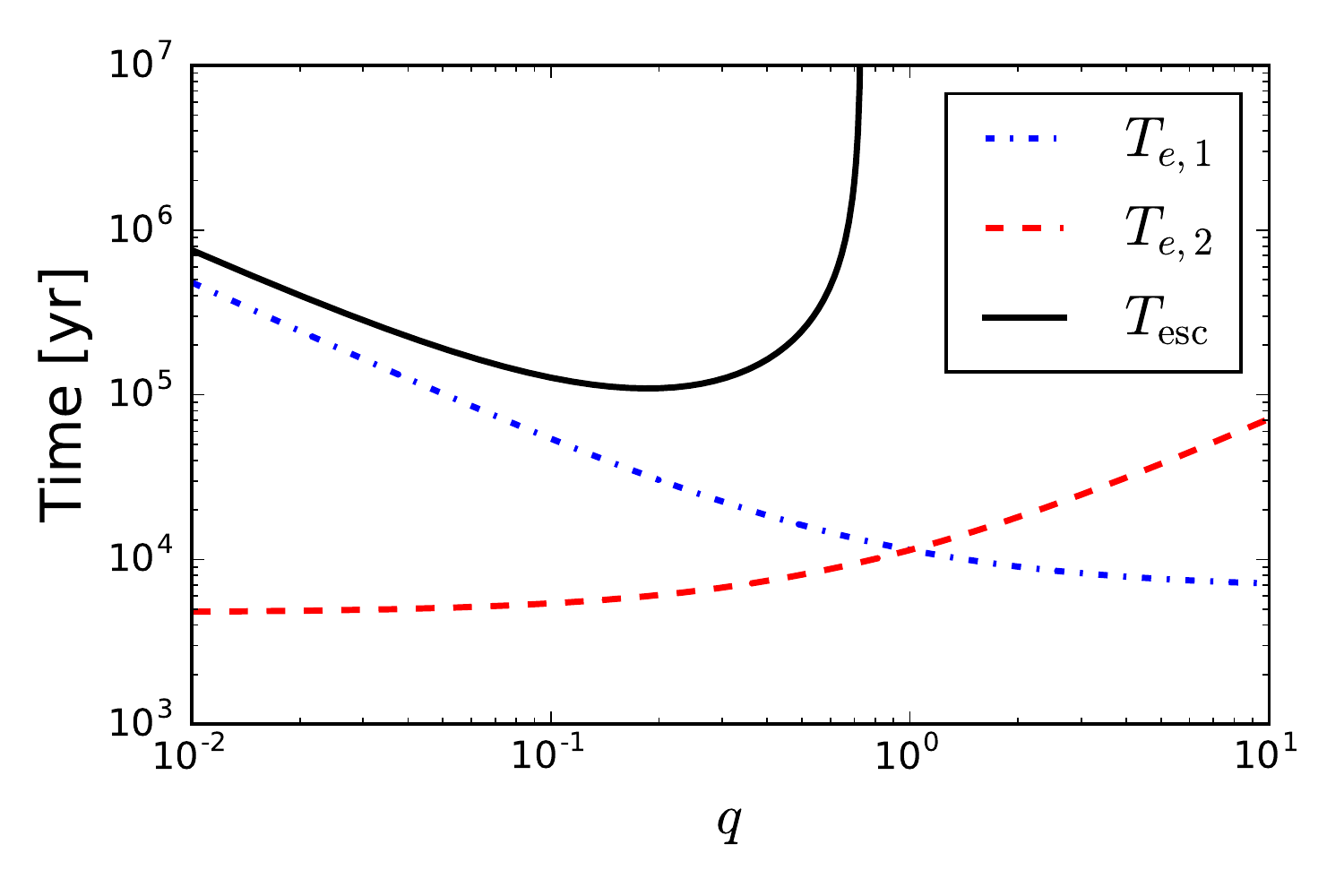}
\caption{Resonance escape timescale $T_{\rm esc}$ as a function of
  $q=m_1/m_2$ for a system with $m_1+m_2=15M_\oplus$,
  $M_\star=1M_\odot$, and $T_{e,i}=(M_\oplus/m_i)10^5\,{\rm yr}$. In
  this calculation we consider $j=5$ (the 3:5 resonance) and
  $T_m=1$~Myr, although the result is not sensitive to these
  parameters. The eccentricity damping time $T_{e,i}$ is also shown
  for reference. We see that $T_{\rm esc}\gtrsim T_{e,i}$ for all
  values of $q$, and near $q_{\rm crit}$ the difference between
  $T_{e,i}$ and $T_{\rm esc}$ becomes significant.}
\label{stability_esc}
\end{figure}

We end this section by emphasizing that, even for an overstable
equilibrium, the resonance escape time $T_{\rm esc}=\tau_{\rm esc}T_0$
can be much larger than the eccentricity damping time $T_{e,i}$ when
the mass ratio is close to the stability boundary $q_{\rm crit}$. Thus, a nontrivial 
portion of such systems may stay close to
resonance until migration ends, even when $T_{e,i}$ is much smaller
than the disk lifetime. Figure \ref{stability_esc} gives an example
for the 3:5 resonance.

\section{N-Body Simulations}

In the previous sections, we have studied the dynamics of a two-planet
system near resonance analytically using a Hamiltonian formalism. N-body
calculations can be used to verify the results obtained from this
Hamiltonian approach. More importantly, N-body calculations are
necessary in order to study the migration of planet between resonances
and compare the timescales the planets spend inside and outside the
resonance.

In this section, we use the \textit{MERCURY} code \citep{Chambers99} to
integrate the evolution of a two-planet system, with planet-disk
interaction modeled by adding an extra force on each planet (e.g., \citealt{Snellgrove01,NelsonPapaloizou02,LeePeale02})
\eq{
\left.\frac{d\mathbf v_i}{dt} \right|_{\rm disk}= -\frac{2\dot{\mathbf r}_i\cdot\hat{\mathbf r}_i}{T_{e,i}}\hat{\mathbf r}_i-\frac{\dot{\mathbf r}_i\cdot\hat{\pmb \varphi}_i}{2T_{m,i}}\hat{\pmb \varphi}_i,
}
where $\hat {\mathbf r}_i$ and $\hat {\pmb \varphi}_i$ are the unit
vectors in radial and azimuthal directions, $T_{e,i}$
and $T_{m,i}$ are the eccentricity damping and migration timescales as
previously defined [Eqs. \eqref{diss_1}-\eqref{diss_2}], and this
model of planet-disk interaction corresponds to $p_i=2$. We always use
$e_1=e_2=0$ as the initial condition far from resonance. Non-resonant
interaction between the planets could excite finite eccentricities
before entering resonance to allow capture.

\subsection{Stable resonance capture of similar mass planets}

\begin{figure}
\centering
\includegraphics[width=\columnwidth]{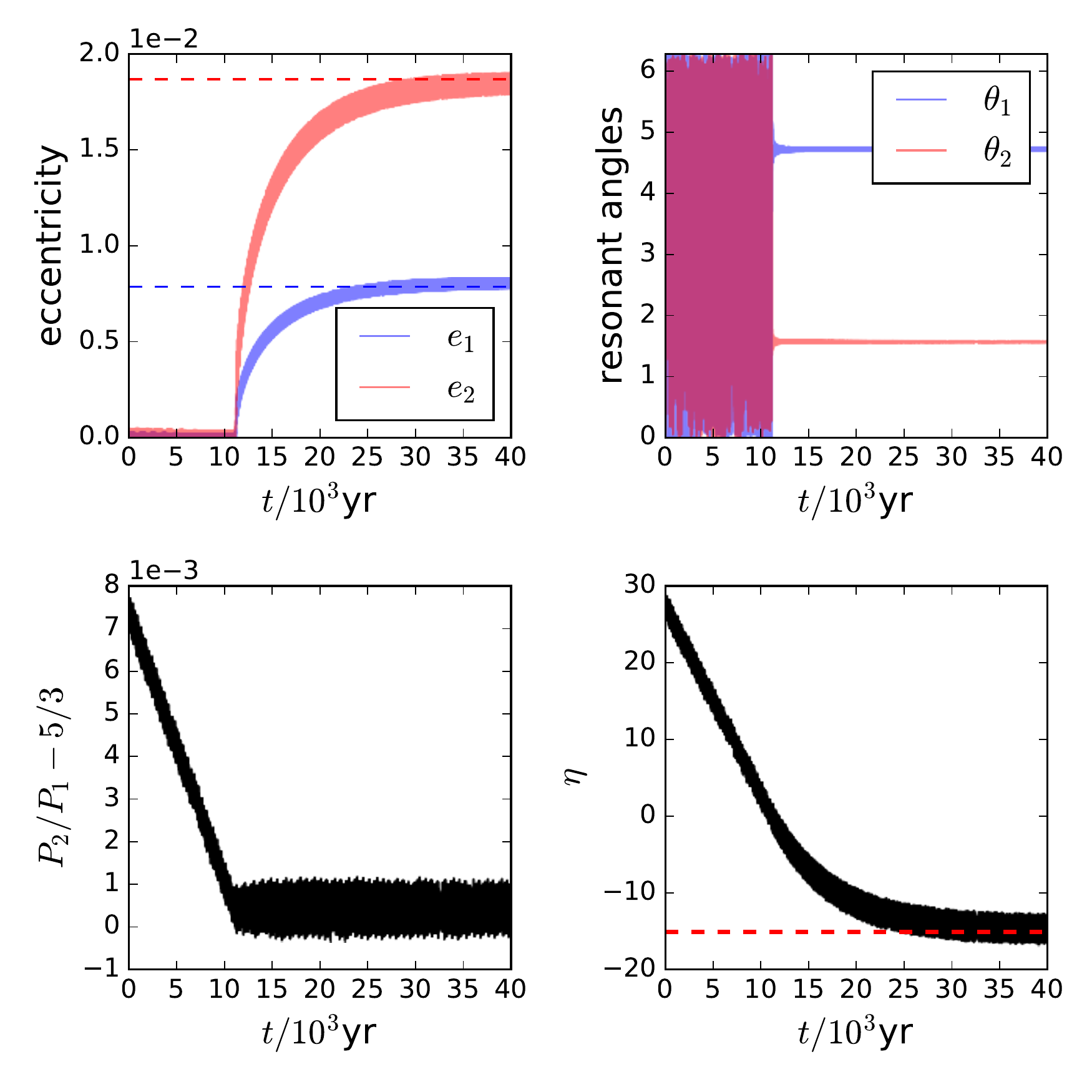}
\caption{$N$-body simulation of capture into the 3:5 MMR of a system with the same 
parameters as in Figure \ref{simMassStb}. In the upper left panel, 
the analytical prediction of the final planet eccentricity (see Appendix A for the calculation) 
is plotted as blue/red-dashed line. In the lower right panel, 
the analytical prediction of the final (equilibrium) $\eta$ is shown as red-dashed line. 
These analytical results agree with the N-body results very well; the 
qualitative behaviors of the resonance angles (upper right panel) and period ratio 
(lower left panel) also match those in Figure \ref{simMassStb}. 
}
\label{numerics1}
\end{figure}

First we use N-body integrations to test the validity of our
analytical results for the case of stable resonance capture. Figure
\ref{numerics1} shows the evolution of a system with the same
parameter as in Figure \ref{simMassStb}. Note that in our $N$-body
integrations, we have chosen the initial condition (initial
$\alpha=a_1/a_2$) further away from resonance (than in
Fig.~\ref{simMassStb}) in order not to miss any
nontrivial behavior of the system before entering the resonance. We
see that the evolution of the system agrees with our previous
results. In particular, the equilibrium eccentricities of the planets and the
equilibrium $\eta$ match the analytical prediction very well. We
conclude that our analytical Hamiltonian approach captures the
relevant dynamics of the resonance when the eccentricities of the
planets are not too large.

It is worth noting that the system we choose for Figure
\ref{numerics1} has a similar total planet mass and semi-major axes as
the five observed pairs of super Earths close to the second-order MMR
(see Section 1). Further N-body examinations show that capture is
still likely even when the planet masses are reduced to
$m_1=m_2=2.5M_\oplus$ (while keeping the dissipation timescales at
$T_e=10~{\rm kyr}, T_{m,1}=4~{\rm Myr}, T_{m,2}=2~{\rm Myr}$). This
suggests that capture into second-order resonance should be common
even for planets as small as several earth masses if the planets
undergo convergent migration in a disk with migration time
$T_{m,i}\gtrsim 1\,{\rm Myr}$.

\subsection{Overstable resonance capture and escape}

\begin{figure}
\centering
\includegraphics[width=\columnwidth]{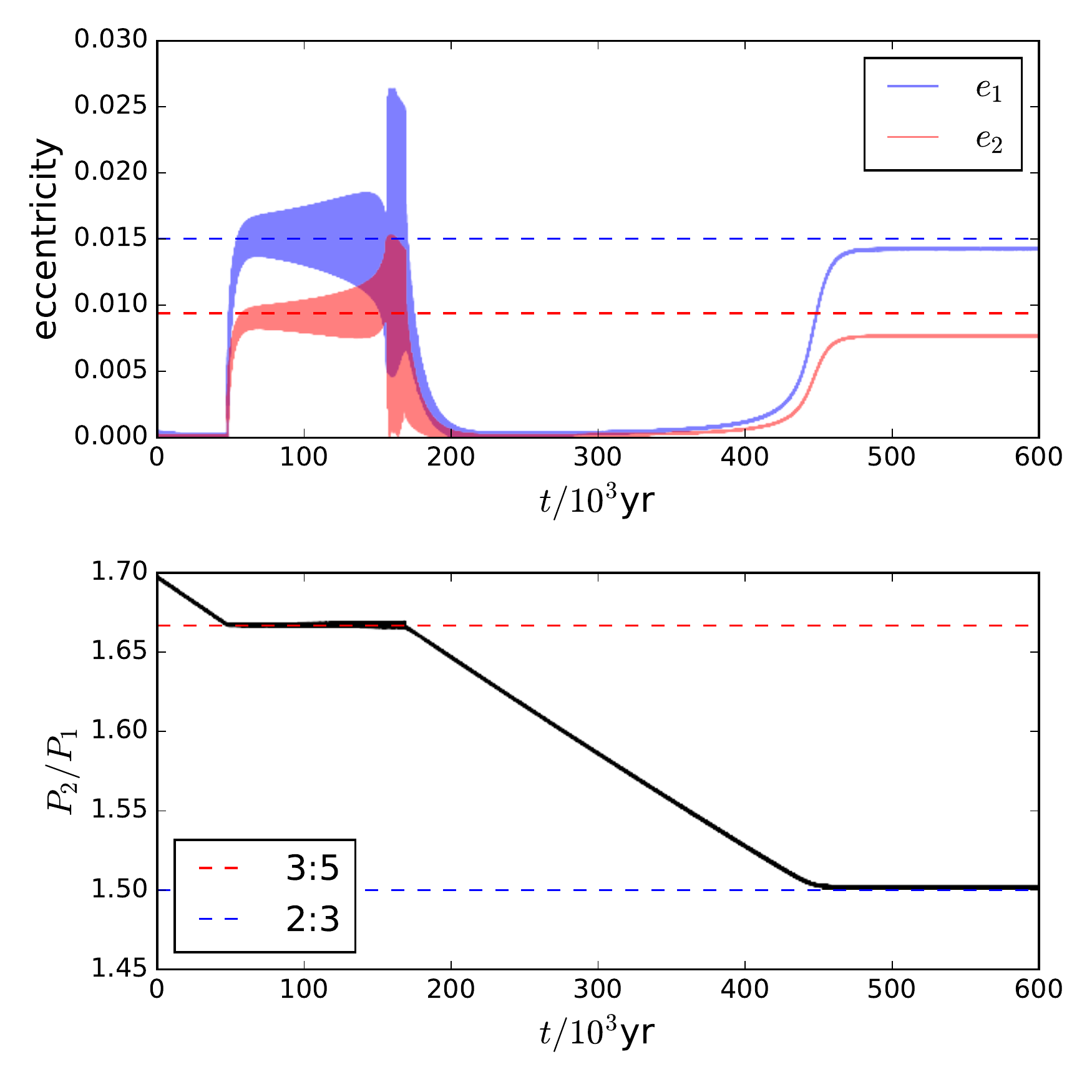}
\caption{$N$-body simulation of capture into and escape from 3:5 MMR of a
  system with the same parameters as in Figure
  \ref{simMassOverstb1}. In the eccentricity panel (left), the theoretical
  predictions of equilibrium eccentricities for the 3:5 MMR are plotted in
  blue/red-dashed lines. The system's behavior qualitatively agrees
  with that in Figure \ref{simMassOverstb1}, while the escape
  timescale differs by about a factor of 2; the difference may come
  from the fact that the actual system (i.e. the one in this figure) has a
  larger eccentricity when entering the resonance. After escaping from
  the 3:5 resonance, the system is subsequently captured into 2:3 mean
  motion resonance (this time the capture is stable). Note that the
  time the planet spends traveling between the two resonances is not
  much longer than the escape time.}
\label{numerics2}
\end{figure}

Next we consider the case of overstable resonance capture and the
escape timescale from the resonance. Figure \ref{numerics2} shows the
$N$-body simulation result for the evolution of a system with the same parameters as Figure
\ref{simMassOverstb1}. In Figure \ref{simMassOverstb1} we can already
observe that the actual escape time from the 3:5 resonance (i.e. the
length of time when the system is in resonance, in this case $\simeq
240\,{\rm kyr}$) is significantly longer than the eccentricity damping
timescale ($T_e=10~{\rm kyr}$); this trend is also reflected in the
estimation of $T_{\rm esc}$ based on eigenvalue of the linearized
system (see Section 5.4), which gives $T_{\rm esc}
\simeq 55\,{\rm kyr}$. Our $N$-body calculation (Fig.~\ref{numerics2}) also shows a large
$T_{\rm esc}\simeq 140\,{\rm kyr}$ for the same system. The difference
between the $N$-body result and the result of Figure
\ref{simMassOverstb1} may come from the fact that in the $N$-body
integration, the planets have larger eccentricities when entering the
resonance.
\textbf{This difference may also come from the fact that our analytical estimation calculates $T_{\rm esc}$ based on the growth rate of libration amplitude for small libration around the fixed point; when the amplitude is larger the growth rate may be different.}
Thus, although our intuitive prediction that $T_{\rm
  esc}\sim T_e$ is roughly correct, the actual escape time $T_{\rm
  esc}$ is very often larger than $T_e$ by more than an order of
magnitude. A direct consequence of this is that the system can spend relatively
long time in resonance compared to the time between resonances even
though $T_e\ll T_m$. For instance, in Figure \ref{numerics2} we see
that the system spends roughly equal time in the 3:5 resonance and
between the 3:5 and 2:3 resonances, while $T_e/T_m$ is only
$1/400$. Thus, in the absence of other physical effects not considered in
our analysis (see Section 7.2), we should expect more overstable systems in
MMR than observed, since there is a nontrivial probability for such system to
end its migration before exiting resonance. (Note that these systems
become stable once the dissipative perturbations, i.e. migration
torques, are gone.)

\section{Summary and Discussion}

\subsection{Summary of results}

Motivated by the observations of {\it Kepler} multi-planet systems
(see Section 1), we have carried out a theoretical study on the
dynamics of second-order mean motion resonances (MMRs) for planets
migrating in protoplanetary disks. In particular, we have examined the
mechanism and the conditions of capturing the migrating planets into a
second-order MMR, and studied the stability of the captured state, in
order to determine whether and on what timescale the planets may
escape from the resonance.  We confirm our theoretical calculations with
numerical $N$-body simulations.  Our key results can be summarized as
follows.

\begin{enumerate}

\item When one of the planets has a negligible mass compared to the
  other, the Hamiltonian of the planets near a second-order MMR can be
  reduced to that of a one degree of freedom system. In this limiting
  case (``restricted three-body problem''), the dynamics of MMR
  capture and its stability property can be explicitly calculated (Sections
  2-3).  When the planets have comparable masses, such simplification
  is no longer possible and the Hamiltonian has two degrees of freedom
  (Section 5). Nevertheless, the dynamics of MMR capture and the stability
  share similar features as the restricted three-body problem.

\item Planets can be captured into a second-order MMR only when the
  migration is convergent (Sections 2.2 and 5.2).
  For divergent migration, there is only a transient
  eccentricity excitation when the system passes the resonance,
  with the maximum eccentricity of the planet reaching
  $e_{1,2}\sim\sqrt{\mu_{2,1}}$ (where $\mu_i=m_i/M_\star$, and $m_{1,2}$ and $M_\star$ 
  are the masses of the planets and the host star, respectively).

\item To capture the planets into a second-order MMR, the migration
  timescale $T_m$, the eccentricity damping timescale $T_e$ due to
  planet-disk interaction, and the initial (``pre-resonance'')
  eccentricity must satisfy Eq.~(\ref{trap_cond_1}).
  These conditions are distinct from those required for the
  first-order MMR capture (Appendix B, available online).  When the migration timescale is shorter
  than that required by Eq.~(\ref{trap_cond_1}),
  the second-order resonance capture becomes probabilistic, and the
  capture probability decreases with decreasing $T_m$.
  In general, capture is easier for more massive planets and longer
  migration and eccentricity damping timescales.  For planets in a
  disk with a migration timescale of $T_m\sim 1$~Myr, these conditions can
  be satisfied for super-Earths ($m_i\gtrsim$ a few $M_\oplus$) at
  $a\sim 0.1$~AU (see Fig.~\ref{trapRegion}).

\item Following the resonance capture of the planets, an equilibrium state can be reached
  in which eccentricity excitation due to resonant planet-planet
  forcing balances eccentricity damping due to planet-disk
  interaction. However, this equilibrium may be overstable, leading to
  partial or permanent escape of the planets from the resonance. For the
  restricted three-body problem, we find that the captured equilibrium
  state is always overstable (with the growth time of order $T_e$)
  when $q\equiv m_1/m_2 \ll 1$ ($q$ is the mass ratio of the inner and outer
  planets), and stable when $q\gg 1$ (Sections 2.4 and 3.3).
  For comparable mass planets, the stability of the equilibrium state
  depends not only on $q$, but also on the ratio of
  the eccentricity damping rate $q_e=T_{e,2}/T_{e,1}$ and the specific
  resonance (Section 5.4). Thus, in general, for planets captured into
  a second-order MMR, there are three possible outcomes:

\begin{enumerate}
\item When $q\gtrsim q_e^{1/2}$ ($q\gtrsim 2q_e^{2/3}$ for 1:3 MMR),
  the planets are stably (permanently) captured with their period
  ratio held close to the resonance and the resonant angles librating
  (see Figs.~\ref{toyModelStb}, \ref{simMassStb} and \ref{numerics1}).
\item When $q\lesssim q_e^{1/2}$ ($q\lesssim 2q_e^{2/3}$ for 1:3 MMR)
    and $|\eta_{\rm eq}|\lesssim 1$ [where $\eta_{\rm eq}$ is the value
    of the resonance parameter at the equilibrium; see
    Eq.~\eqref{eta_eq_scale} for a general estimate
    ], the planets
    have circulating resonant angles but 
    only partially escape from the resonance:
    the period ratio is still held close to and
    oscillate around the resonant value (see
    Figs.~\ref{toyModelOverstb2} and \ref{simMassOverstb2}).  Note that
    the second condition is met only when the planets are very massive
    ($\mu_0\gtrsim 10^{-2}$; see Eq.~\ref{eq:muodefine})
    or the difference between $T_{m,1}$ and $T_{m,2}$ is very small.
\item When $q\lesssim q_e^{1/2}$ ($q\lesssim 2q_e^{2/3}$ for 1:3 MMR)
  and $|\eta_{\rm eq}|\gtrsim 1$, the planets escape from the resonance
  and the period ratio no longer stays close to the resonant value (see
  Figs.~\ref{toyModelOverstb1}, \ref{simMassOverstb1} and \ref{numerics2}).
\end{enumerate}
{\citet{Delisle15} have derived a criterion for stable capture in terms of the (equilibrium) eccentricity ratio of the planets. Our result agrees with theirs when the relationship between the eccentricity ratio and the mass ratio is spelled out (see Figure \ref{eccRatio}).  Our calculation shows that in terms of $q$ and $q_e$, the criterion for stable capture for the 1:3 MMR is different from that for the other MMRs: The 1:3 resonance has a larger parameter space for unstable capture.}

\item For planets that are captured but eventually leave the resonance,
 the linear growth time of the overstability is of order $T_e$,
  but can become much larger for systems near the stability threshold 
  (see Figs.~\ref{stability_te}, \ref{stability_qe} and
  \ref{stability_esc}).  These analytical results and our $N$-body
  simulations both show that the escape time from resonance for these
  planets can be $10$-$100$ times larger than $T_e$.  As a result, the
  time these planets spend in the resonance can be comparable to the
  time they spend migrating between resonances (Section 6.2 and 
Fig.~\ref{numerics2}).

\end{enumerate}

\subsection{Discussions}

A major goal of this paper is to evaluate the probability of
forming planet pairs in second-order MMRs by resonance capture during
planet migration. There is a common preconception that capture into
second-order MMRs is difficult because the planet perturbation
associated with a second-order resonance may be too weak to counter
the eccentricity damping due to planet-disk interaction. Our analysis
of the conditions for capture indicates that capturing the planets
into second-order MMRs is entirely possible, and only moderately large
$T_e$ and $T_m$ are required. Therefore, a significant fraction of
planets formed in gaseous disks could have been (permanently or
temporarily) captured into second-order MMRs during
migration. However, our analysis of the stability of the ``captured" equilibrium
shows that such capture is very often overstable: Following a
capture, the system oscillates around the equilibrium state with
growing amplitude, and eventually escapes from the resonance.  For
realistic systems, capture is stable [case 4(a) above] only when
$q\gtrsim 1$, while convergent migration, which is necessary for
capture in the first place, usually requires $q\lesssim 1$ for
disk-driven Type I migration (see the end of Section 5.4 for
discussion). We suggest that this is a likely reason why the observed
period ratio distribution for {\it Kepler} multi-planet systems barely
shows any peak near second-order MMRs (see Section 1). In particular,
{\it it is much more difficult to find a system stably captured into a
second-order MMR than into a first-order MMR}, because for first-order
MMRs there is a large region in the parameter space where capture can be
stable for $q\lesssim 1$ [see Appendix B and \citet{DeckBatygin15}].

Our results can be used to explain some of the observed features of
multi-planet systems in or near second-order MMRs, and comparison
between our results and observations poses several interesting
questions. As noted in Section 1,
the {\it Kepler} sample currently contains 
at least five pairs of planets with 
period ratio within $7\times 10^{-3}$ of the exact resonance
(and within $4\times 10^{-5}$ for two of the systems),
and most of them are low-mass ($\sim 10M_\oplus$) planets, and have mass ratio $q\sim 1$. 
Our analysis suggests that it is indeed quite possible for these
planets to be captured into resonance during migration, and the mass
ratio could be explained by the fact that convergent migration and
stable capture can be simultaneously satisfied only for $q\sim 1$.
The low masses of these pairs could be explained by the fact that
planets with larger masses and $q\sim 1$ captured into a second-order
MMR can become dynamically unstable, especially for MMR with 
$j>5$ (such as $5:7$, $7:9$) 
[\citet{AndrePapaloizou16} provide an example of such dynamical instability].

As noted before, the (de-biased) period ratio distribution of {\it Kepler}
multi-planets exhibits no significant features near second-order
MMRs except for the $3:5$ MMR [see Fig. 20 of \citet{SteffenHwang15}]. 
The overall paucity of planets in second-order MMRs
(compared to those in first-order MMRs) 
\footnote{We note that from period ratio alone, it can be difficult to
  infer whether a pair of planets is in a second-order MMR, because
  the width of the resonance is of order $\mu$, while the detuning of
  the equilibrium state from the ``exact" resonance is of order
  $T_e/T_m$ and is typically $\gtrsim \mu$. By contrast, for a
  first-order MMR, the detuning is also of order $T_e/T_m$, but the
  resonance width is $\sim \mu^{2/3}$.}  can be explained by the fact
that the captured planets in second-order MMRs are less stable.  The
fact that the $3:5$ MMR is more significant than the others could be
caused by two effects: (i) For the $1:3$ MMR, the difference in
stability criteria [see 4(a)-(c) in Section 7.1] implies that fewer
planets can remain in this resonance compared to the 3:5 MMR; (ii) for
resonances with $j>5$ (e.g. $5:7$, $7:9$), the planet's spacing is so
small that that few planets reach these resonances before being
captured into another resonance or becoming dynamically unstable.

There are several factors we did not consider in this paper and they may
impact the population of planets in second-order MMRs. First, the
planet migration model we used assumes that migration is
one-directional and the planet period ratio varies smoothly. Migration
can shift from inward to outward (and sometimes convergent to
divergent) as the disk evolves and disperses (e.g., \citealt{ChatterjeeFord15}); 
this might move planets initially captured by convergent
migration out of resonance. Second, waves and turbulences in
protoplanetary disks may strongly affect second-order MMRs, since the
resonant interaction is relatively weak. \citet{AndrePapaloizou16}
show that the waves in the disk produced by the planets do not affect
capture into second-order MMR in some scenarios, while destabilizes
the system in others. Third, we assumed that before migrating, the
planets have accreted most of (at least a significant fraction of)
their masses. It is possible that planets accrete most of their masses
when the disk is mostly dispersed and migration is too slow for a
significant fraction of planets to travel far enough to encounter a
resonance. Finally, when passing a second-order MMR, in principle it
is possible that inclination will be excited as well as (or instead
of) eccentricity excitation. In this case, the increased inclination
of captured planets leads to a lower probability of observing both
planets, making the
population of planets in second-order MMRs invisible to transit
surveys.  This effect will be studied in our next paper.

\section*{Acknowledgements}
This work has been supported in part by NASA grants NNX14AG94G and
NNX14AP31G, and a Simons Fellowship from the Simons Foundation.  WX
acknowledges the supports from the Hunter R. Rawlings III Cornell
Presidential Research Scholar Program and Hopkins Foundation 
Research Program for Undergraduates.

\bibliographystyle{mnras}
\bibliography{BIB}

\onecolumn
\appendix
\section{Dynamics of similar mass planets near second-order mean motion resonance}

\subsection{Derivation of the reduced Hamiltonian}

The Hamiltonian of two massive planets (with masses $m_1$ and $m_2$) orbiting their host 
star (with mass $M_\star$) can be written as
\eq{
H= H_0+H_1,
}
with
\eal{
H_0 &=\sum_{i=1,2}\left(
\frac{|\mathbf p_i|^2}{2\tilde m_i}-\frac{G\tilde M_i\tilde m_i}{|\mathbf{r}_i|}\right),\\
H_1 &\simeq-Gm_1m_2\left(\frac{1}{|\mathbf{r}_1-\mathbf {r}_2|}-\frac{\mathbf {r}_1\cdot\mathbf{r}_2}{|\mathbf{r}_2|^3}\right),
}
where $\tilde M_i\simeq M_\star$ and $\tilde m_i\simeq m_i$ are the
Jacobi masses. In the following calculation we ignore the difference
between the Jacobi and physical masses; the error introduced by this
simplification is negligible. With this approximation, we have
\eq{
H_0 = -\sum_{i=1,2}\frac{G^2M_\star^2m_i^3}{2\Lambda_i^2},
}
where $\Lambda_i \equiv m_i\sqrt{GM_\star a_i}$ is conjugate to the mean longitude $\lambda_i$.

We consider the system near the $j-2:j$ resonance.  The perturbation
$H_1$, after truncating the terms of 
higher order than $e^4, \mu_i e^2$ (where $\mu_i=m_i/M_\star$)
and averaging over non-resonant angles, can be written as
\eq{
H_1 =& -\frac{G^2M_\star m_1 m_2^3}{\Lambda_2^2}\left[f_{0,1}+f_{0,2}\left(\frac{2\Gamma_ 1}{\Lambda_1}+\frac{2\Gamma_ 2}{\Lambda_2}\right)+f_{0,10}\sqrt{\frac{4\Gamma_ 1\Gamma_ 2}{\Lambda_1\Lambda_2}}\cos(\gamma_ 1-\gamma_ 2)+f_{j,45}\frac{2\Gamma_ 1}{\Lambda_1}\cos\Bigl(j\lambda_2-(j-2)\lambda_1+2\gamma_ 1\Bigr)\right.\\
&\left.+f_{j,49}\sqrt{\frac{4\Gamma_ 1\Gamma_ 2}{\Lambda_1\Lambda_2}}\cos\Bigl(j\lambda_2-(j-2)\lambda_1+\gamma_ 1+\gamma_ 2\Bigr)+f_{j,53}\frac{2\Gamma_ 2}{\Lambda_2}\cos\Bigl(j\lambda_2-(j-2)\lambda_1+2\gamma_ 2\Bigr)\right].
}
Here $\Gamma_ i \equiv \Lambda_i\left(1-\sqrt{1-e_i^2}\right)$ is
conjugate to $\gamma_ i \equiv-\varpi_i$, and $f_{m,n}$ are functions
of $\alpha=a_1/a_2$. The expression for $f_{m,n}$ can be found in
Appendix B of \citet{MurrayDermott99}. Note that $f_{m,n}$ refers to
$f_n$ with $j=m$ in the book, with the exception that when $j=3$,
$f_{j,53}=f_{53}-3^{7/3}/8$ (this extra piece comes from an ``indirect
term", which is present only for the 1:3 resonance.)

To simplify the Hamiltonian, we introduce the following sets of canonical variables:
\eal{
& \Phi_1 = \frac{j-2}{j}\Lambda_2+\Lambda_1, &&\phi_1 = \lambda_1\\
&\Phi_2 = \frac 1j \Lambda_2 -\frac 12 \Gamma_ 1-\frac 12 \Gamma_ 2 && \phi_2 = j\lambda_2-(j-2)\lambda_1\\
& \Psi_1 = \Gamma_ 1, &&\psi_1 = \gamma_ 1+\frac 12 \phi_2\\
& \Psi_2 = \Gamma_ 2, &&\psi_2 = \gamma_ 2+\frac 12 \phi_2
}
Since $\phi_1$ and $\phi_2$ do not appear in the Hamiltonian, $\Phi_1$ and $\Phi_2$ are constants of motion.

Next we expand the Hamiltonian around the $j-2:j$ resonance, which is given by 
\eq{\left(\frac{n_2}{n_1}\right)_{\rm res} = \alpha_{\rm res}^{3/2} = \frac{(j-2)}{j}.}
In the following calculation, the $f$ parameters are all evaluated at
$\alpha=\alpha_{\rm res}$. The Keplerian part of the Hamiltonian is (keeping terms up
to order $e^4$ and $\mu_ie^2$)
\eq{
H_0 = -\frac{3j^2}{8}\left(\alpha_{\rm res}q^{-1}+1\right)\frac{GM_\star^2m_2^3}{\Lambda_{2,0}^4}\left(\Psi_1+\Psi_2+2\Phi_2-\frac{2}{j}\Lambda_{2,0}\right)^2 + \text{const},
}
where for convenience we have defined $q=m_1/m_2$ and
\eq{
\Lambda_{2,0}\equiv(\alpha_{\rm res}+q)^{-1}\alpha_{\rm res}^{-1/2}\Phi_1.
}
Note that $\Lambda_{2,0}\simeq \Lambda_2$ near the resonance, and $\Lambda_{2,0}=\Lambda_2$ when $\alpha=\alpha_{\rm res}$. Similarly, the perturbation part of the Hamiltonian can be written as
\eq{
H_1 =& -\frac{2G^2M_\star m_2^4}{\Lambda_2^3}\left[
-\frac j2 \alpha_{\rm res}(\alpha_{\rm res}+q)\frac{df_{0,1}}{d\alpha}(\Psi_1+\Psi_2)
+f_{0,2}(\alpha_{\rm res}^{-1/2}\Psi_1 + q\Psi_2)
\right.\\
&+f_{0,10}\alpha_{\rm res}^{-1/4}q^{1/2}\sqrt{\Psi_1\Psi_2}\cos(\psi_1-\psi_2)+f_{j,45}\alpha_{\rm res}^{-1/2}\Psi_1\cos(2\psi_1)\\
&+\left.f_{j,49}\alpha_{\rm res}^{-1/4}q^{1/2}\sqrt{\Psi_1\Psi_1}\cos(\psi_1+\psi_2) + f_{j,53}q\Psi_2\cos(2\psi_2)\right] + \text{const}.
}

To further simplify the Hamiltonian we scale it to the dimensionless form
\eq{
\mathcal H \equiv \frac{32}{3j^2}\mu_0^{-1}\frac{\Lambda_{2,0}^4}{GM_\star^2m_2^3}H
}
where $\mu_0\equiv \mu_1+\mu_2\alpha_{\rm res}$, and introduce the scaled canonical variables
\eal{
&x_i = \sqrt{\frac{2\Psi_i}{\mu_1\Lambda_{2,0}}}\cos\psi_i,\\
&y_i = \sqrt{\frac{2\Psi_i}{\mu_1\Lambda_{2,0}}}\sin\psi_i.
}
With this nondimensionalization, time is scaled to
\eq{
\tau=t/T_0,~~~T_0\equiv \left(\frac {3j^2} {32}\mu_0n_2\right)^{-1}.
}
The Hamiltonian can then be written as (ignoring the constant term)
\eq{
-\mathcal H = (x_1^2+x_2^2+y_1^2+y_2^2+\eta_0+c_0)^2 + c_1x_1^2 + c_2x_2^2+c_3y_1^2+c_4y_2^2+c_5x_1x_2+c_6y_1y_2,
}
where
\eq{
\eta_0 = \frac{4}{\mu_0}\left(\frac{\Phi_2}{\Lambda_{2,0}}-\frac{1}{j}\right) = \frac{4}{j\mu_0}\left[1-\left(\frac{\alpha}{\alpha_{\rm res}}\right)^{1/2}\right]-\mu_2^{-1}\alpha_{\rm res}^{1/2}e_1^2-\mu_1^{-1}e_2^2,
}
and
\eal{
&c_0 = -\frac{8}{3j}\frac{df_{0,1}}{d\alpha}\\
&c_1 = \left[\frac{3j^2}{32}(\alpha_{\rm res}+q)\right]^{-1}\alpha_{\rm res}^{-1/2}(f_{0,2}+f_{j,45})\\
&c_2 = \left[\frac{3j^2}{32}(\alpha_{\rm res}+q)\right]^{-1}q(f_{0,2}+f_{j,53})\\
&c_3 = \left[\frac{3j^2}{32}(\alpha_{\rm res}+q)\right]^{-1}\alpha_{\rm res}^{-1/2}(f_{0,2}-f_{j,45})\\
&c_4 = \left[\frac{3j^2}{32}(\alpha_{\rm res}+q)\right]^{-1}q(f_{0,2}-f_{j,53})\\
&c_5 = \left[\frac{3j^2}{32}(\alpha_{\rm res}+q)\right]^{-1}\alpha_{\rm res}^{-1/4}q^{1/2}(f_{0,10}+f_{j,49})\\
&c_6 = \left[\frac{3j^2}{32}(\alpha_{\rm res}+q)\right]^{-1}\alpha_{\rm res}^{-1/4}q^{1/2}(f_{0,10}-f_{j,49}).
}
These coefficients are all constants of order unity.

Finally, we make can make all terms in the Hamiltonian quadratic by writing it as
\eq{\label{ap_Ham}
-\mathcal H = (x_1^2+x_2^2+y_1^2+y_2^2+\eta)^2 + (Ax_1+Bx_2)^2 + (Cy_1+Dy_2)^2 +E^2x_1^2,
}
where $A,B,C,D,E$ and $\eta$ can be found by solving
\eq{
2(\eta-\eta_0-c_0) + A^2 + E^2 &= c_1\\
2(\eta-\eta_0-c_0) + B^2 &= c_2\\
2(\eta-\eta_0-c_0) + C^2 &= c_3\\
2(\eta-\eta_0-c_0) + D^2 &= c_4\\
2AB &= c_5\\
2CD &= c_6
}
Note that if the values of $c_i$ were arbitrary than we could not
guarantee that $E$ is real. However, if we can choose the last term of
\eqref{ap_Ham} between $E^2x_1^2$ and $E^2x_2^2$, then we can always
choose a real $E$ for any $c_i$. For our application, it happens that
$E$ is always real when the last term is $E^2x_1^2$. This result is
obtained by numerically solving for the coefficients for $j=3,5,7,9$
and $q\in[10^{-3},10^3]$.

\subsection{Finding the fixed points}

From the Hamiltonian \eqref{ap_Ham} we can immediately see that the origin is always a fixed point since there is no linear term. Also, thanks to the quadratic form of the Hamiltonian, we notice that when $\eta<0$, 
\eq{
(x_1,x_2,y_1,y_2) = \left(0,0,\pm D\sqrt{\frac{-\eta}{C^2+D^2}},\mp C\sqrt{\frac{-\eta}{C^2+D^2}}\right)
}
are stable fixed points since they are the global maxima of the Hamiltonian.

The other fixed points need to be solved by considering the equations of motion:
\eal{
\label{ap_eom_1}\dot x_1 &= -\frac{\partial \mathcal H}{\partial y_1} = 4y_1(x_1^2+x_2^2+y_1^2+y_2^2+\eta) + 2C(Cy_1+Dy_2)\\
\label{ap_eom_2}\dot x_2 &= -\frac{\partial \mathcal H}{\partial y_2}= 4y_2(x_1^2+x_2^2+y_1^2+y_2^2+\eta) + 2D(Cy_1+Dy_2)\\
\label{ap_eom_3}\dot y_1 &= \frac{\partial \mathcal H}{\partial x_1}= -4x_1(x_1^2+x_2^2+y_1^2+y_2^2+\eta) - 2A(Ax_1+Bx_2) - 2E^2x_1\\
\label{ap_eom_4}\dot y_2 &= \frac{\partial \mathcal H}{\partial x_2}= -4x_2(x_1^2+x_2^2+y_1^2+y_2^2+\eta) - 2B(Ax_1+Bx_2).
}
The fixed points correspond to $(x_1,x_2,y_1,y_2)$ for which the RHS of all four equations are zero. From the first two equations, it is easy to see that $y_1,y_2$ are either both zero or both nonzero. When $y_1,y_2$ are nonzero, Eqs. \eqref{ap_eom_1}-\eqref{ap_eom_2} yield
\eq{
y_1/y_2 = C/D\text{~~~or~~~}-D/C
}
and
\eq{
x_1^2+x_2^2+y_1^2+y_2^2+\eta = -\frac{C^2+D^2}{2} \text{~~~or~~~} 0
}
respectively. Similarly, for Eqs. \eqref{ap_eom_3}-\eqref{ap_eom_4}, we see that $x_1,x_2$ are either both zero, or satisfy
\eq{
\frac {x_1} {x_2} = \frac{A^2-B^2+E^2\pm \sqrt{(A^2-B^2+E^2)^2+4A^2B^2}}{2AB}
}
and
\eq{
x_1^2+x_2^2+y_1^2+y_2^2+\eta = -\frac 1 4\left(A^2+B^2+E^2\pm \sqrt{(A^2-B^2+E^2)^2+4A^2B^2}\right).
}
When $x_1,x_2,y_1,y_2$ are all nonzero, the two conditions for $x_1^2+x_2^2+y_1^2+y_2^2+\eta$ in general contradict each other. Therefore, we need $x_1,x_2=0$ or $y_1,y_2=0$. Thus, there are four pairs of fixed points beside the origin, and they are
\eal{
&{\rm FP}_1: ~~~(0,0,\pm y_{11}\sqrt{-\eta},\pm y_{21}\sqrt{-\eta})&&\text{when~~}\eta<0\\
&{\rm FP}_2: ~~~(0,0,\pm y_{12}\sqrt{-\eta+\eta_2},\mp y_{22}\sqrt{-\eta+\eta_2})&&\text{when~~}\eta<\eta_2\\
&{\rm FP}_3: ~~~(\pm x_{11}\sqrt{-\eta+\eta_3},\pm x_{21}\sqrt{-\eta+\eta_3},0,0)&&\text{when~~}\eta<\eta_3\\
&{\rm FP}_4: ~~~(\pm x_{12}\sqrt{-\eta+\eta_4},\pm x_{22}\sqrt{-\eta+\eta_4},0,0)&&\text{when~~}\eta<\eta_4
}
Here
\eal{
\eta_2 &= -\frac{C^2+D^2}{2}\\
\eta_3 &= -\frac 1 4\left(A^2+B^2+E^2- \sqrt{(A^2-B^2+E^2)^2+4A^2B^2}\right)\\
\eta_4 &= -\frac 1 4\left(A^2+B^2+E^2+ \sqrt{(A^2-B^2+E^2)^2+4A^2B^2}\right),
}
and
\eq{
y_{11} &= D/\sqrt{C^2+D^2},~~~y_{21} = -C/\sqrt{C^2+D^2}\\
y_{12} &= C/\sqrt{C^2+D^2},~~~y_{22} = D/\sqrt{C^2+D^2}\\
x_{11},x_{12} &= \frac{A^2-B^2+E^2\mp \sqrt{(A^2-B^2+E^2)^2+4A^2B^2}}{A^2-B^2+E^2+2AB\mp \sqrt{(A^2-B^2+E^2)^2+4A^2B^2}}\\
x_{21},x_{22} &= \frac{2AB}{A^2-B^2+E^2+2AB\mp \sqrt{(A^2-B^2+E^2)^2+4A^2B^2}}.
}
Clearly $\eta_4<\eta_3$ and $\eta_{2,3,4}<0$. It is worth noting that the locations of the fixed points depend on $\eta$; such dependence, however, does not affect the stability since the ratio between $x_1,x_2,y_1$ and $y_2$ is fixed for each branch of fixed points.

When planet-disk interactions are included, we introduce extra terms into the equations of motion, and the locations of fixed points (or equilibrium points) change. The equations of motion, including planet-disk interactions, are
\eal{\label{ap_eom_diss_1}
\dot x_1 &= -\frac{\partial \mathcal H}{\partial y_1} - \frac{x_1}{\tau_{e,1}}\\
\dot x_2 &= -\frac{\partial \mathcal H}{\partial y_2} - \frac{x_2}{\tau_{e,2}}\\
\dot y_1 &= \frac{\partial \mathcal H}{\partial x_1} - \frac{y_1}{\tau_{e,1}}\\
\dot y_2 &= \frac{\partial \mathcal H}{\partial x_2} - \frac{y_2}{\tau_{e,2}}\\
\label{ap_eom_diss_2}
\dot \eta &=-\frac{2}{j\mu_0\tau_m}+\frac{x_1^2+y_1^2}{\tau_{e,1}}\left[2+\frac{2p_1\alpha_{\rm res}^{-1/2}}{j(q+\alpha_{\rm res})}\right]+\frac{x_2^2+y_2^2}{\tau_{e,2}}\left[2-\frac{2p_2q}{j(q+\alpha_{\rm res})}\right]
}
Note that in this case $\eta$ is no longer constant and the fixed points are specified by $(x_1,x_2,y_1,y_2,\eta)$, and we solve the equilibrium points numerically. The equilibrium states we are interested in are those correspond to FP$_1$ in the non-dissipative case; and for weak dissipation, they should be close to FP$_1$.

\subsection{Stability analysis}

To determine stability of a fixed point we linearize the equations of motion near the fixed point and numerically calculate the eigenvalues. For the non-dissipative case, the linearized equations of motion are
\eal{
\delta \dot x_1 &= \sum_{i=1,2}\left(-\frac{\partial^2 \mathcal H}{\partial y_1\partial x_i}\delta x_i-\frac{\partial^2 \mathcal H}{\partial y_1\partial y_i}\delta y_i\right),\\
\delta \dot x_2 &= \sum_{i=1,2}\left(-\frac{\partial^2 \mathcal H}{\partial y_2\partial x_i}\delta x_i-\frac{\partial^2 \mathcal H}{\partial y_2\partial y_i}\delta y_i\right),\\
\delta \dot y_1 &= \sum_{i=1,2}\left(\frac{\partial^2 \mathcal H}{\partial x_1\partial x_i}\delta x_i+\frac{\partial^2 \mathcal H}{\partial x_1\partial y_i}\delta y_i\right),\\
\delta \dot y_2 &= \sum_{i=1,2}\left(\frac{\partial^2 \mathcal H}{\partial x_2\partial x_i}\delta x_i+\frac{\partial^2 \mathcal H}{\partial x_2\partial y_i}\delta y_i\right),
}
where the derivatives are evaluated at the fixed point. The eigenvalues can be calculated numerically for a specific $j$ and $q$. Our calculation shows that for $q\in[10^{-3},10^3]$, the stability is independent of $q$: FP$_2$, FP$_3$ and FP$_4$ are always unstable, FP$_1$ is always stable, and FP$_0$ (the origin) is stable when $\eta<\eta_4$ or $\eta>0$, and unstable otherwise.

When planet-disk interactions are included, we similarly linearize the equations of motion \eqref{ap_eom_diss_1} - \eqref{ap_eom_diss_2} near a fixed (equilibrium) point:
\eal{
\delta \dot x_1 &= \sum_{i=1,2}\left(-\frac{\partial^2 \mathcal H}{\partial y_1\partial x_i}\delta x_i-\frac{\partial^2 \mathcal H}{\partial y_1\partial y_i}\delta y_i\right)-\frac{\delta x_1}{\tau_{e,1}}\\
\delta \dot x_2 &= \sum_{i=1,2}\left(-\frac{\partial^2 \mathcal H}{\partial y_2\partial x_i}\delta x_i-\frac{\partial^2 \mathcal H}{\partial y_2\partial y_i}\delta y_i\right)-\frac{\delta x_2}{\tau_{e,2}}\\
\delta \dot y_1 &= \sum_{i=1,2}\left(\frac{\partial^2 \mathcal H}{\partial x_1\partial x_i}\delta x_i+\frac{\partial^2 \mathcal H}{\partial x_1\partial y_i}\delta y_i\right)-\frac{\delta y_1}{\tau_{e,1}}\\
\delta \dot y_2 &= \sum_{i=1,2}\left(\frac{\partial^2 \mathcal H}{\partial x_2\partial x_i}\delta x_i+\frac{\partial^2 \mathcal H}{\partial x_2\partial y_i}\delta y_i\right)-\frac{\delta y_2}{\tau_{e,2}}\\
\delta\dot\eta &= \frac{2x_{1,\rm eq}\delta x_1+2y_{1,\rm eq}\delta y_1}{\tau_{e,1}}\left[2+\frac{2p_1\alpha_{\rm res}^{-1/2}}{j(q+\alpha_{\rm res})}\right]+\frac{2x_{2,\rm eq}\delta x_2+2y_{2,\rm eq}\delta y_2}{\tau_{e,2}}\left[2-\frac{2p_2q}{j(q+\alpha_{\rm res})}\right].
}
The derivatives are all evaluated at the fixed point. Then we can numerically calculate the eigenvalues and determine the stability. In general the eigenvalues only depend on $j,q,\tau_{e,i}$ and $\mu_0\tau_m$ (note that the equilibrium values of $x_i,y_i$ and $\eta$ depend on $\mu_0\tau_m$). How the stability, as well as the escape timescale for the overstable equilibrium point, is affected by these parameters is discussed in Section 5.4.


\section{Capture and Stability of first-order Mean Motion Resonance}

In this section we summarize the dynamics of capture into first-order
MMR and its stability. This serves to illustrate the similarities and
differences between first-order and second-order MMRs.

For two planets in the $j:j+1$ MMR with the inner planet being
massless ($m_1\ll m_2$), the reduced Hamiltonian can be written as
\eq{
-\mathcal H = \eta\Theta+\Theta^2-\sqrt{\Theta}\cos\theta,
}
where
\eal{
&\theta=(j+1)\lambda_2-j\lambda_1-\varpi_1,\\
&\Theta=\left(\frac{3j^2}{8\beta\mu_2}\right)^{2/3}e_1^2,\\
&\eta=\frac{1}{3^{1/3}}(j\beta\mu_2)^{-2/3}\left[j-(j+1)\alpha_0^{3/2}\right],
}
with $\beta=-\alpha f_d(\alpha)\simeq 0.8j$, $\alpha=a_1/a_2$, and
\eq{
\alpha_0=\alpha(1+je_1^2).}
The dimensionless time is
\eq{
\tau=3^{1/3}(j\beta\mu_2)^{2/3} n_1 t \equiv  t/T_0.
}
The case when the outer planet is massless is similar. 

The conditions for resonance capture into first-order MMR are well-understood (e.g., 
\citealt{Henrard82,BorderiesGoldreich84,Peale86}).
Resonance capture requires that $\eta$ decrease in time (corresponding
to convergent migration).  The migration must be sufficiently slow
such that $|d\eta/d\tau|\lesssim 1$. In addition, the initial
(``pre-resonance'') value of $\Theta$ must satisfy $\Theta_0\lesssim
1$. These requirements translate into the conditions for the migration time
and initial eccentricity:
\eq{
T_m\gtrsim \mu^{-4/3}n_1^{-1},\quad
e_0\lesssim \mu^{1/3},
\label{condition-1st}}
where $\mu=\mu_1+\mu_2$. This should be contrasted to (\ref{trap_cond_1})
for the second-order MMR. 
We see that the requirement on $T_m$ for the first-order MMR 
($T_m\gtrsim \mu^{-4/3}n_1^{-1}$) is less stringent than that 
for the second-order MMR ($T_m\gtrsim \mu^{-2}n^{-1}$). Heuristically,
this difference arises because in the second-order MMR, the planet-planet interaction
is weaker, and a more gentle migration is needed for resonance capture.
Moreover, there is no requirement for the eccentricity damping time $T_e$ 
in the case of the first-order MMR capture.
When $T_e$ is shorter than the resonant timescale, a first-order MMR still has an equilibrium state with finite eccentricity (i.e. resonance capture can happen), but for a second-order MMR the eccentricity will be held at zero and there is no equilibrium (i.e. resonance capture cannot happen).

Including the dissipative effect of planet-disk interaction (see 
Eqs.~\ref{diss_1}-\ref{diss_2}), the resonance parameter $\eta$ evolves 
according to (for systems with $m_1\ll m_2$)
\eq{
{d\eta\over d\tau}=-{3^{2/3}j\over 2 (j\beta\mu_2)^{2/3}}\left[{T_0\over T_m}-(p_1+2j){T_0 e_1^2\over T_e}
\right].}
Thus, following capture, the system reaches an equilibrium at
\eq{
e_{\rm eq}=\left[{T_e\over (p_1+2 j)T_m}\right]^{1/2}.
}
The stability of this equilibrium was studied by \citet{GoldreichSchlichting14} 
in the limit $m_1\ll m_2$. They found that capture
is stable only when $e_{\rm eq}\sim (T_e/T_m)^{1/2}\lesssim
\mu_2^{1/3}$. Note that in the same limit, second-order resonance is
always overstable. \citet{DeckBatygin15} extended the stability study to general
mass ratio $q=m_1/m_2$ and eccentricity damping rate ratio $q_e=T_{e,2}/T_{e,1}$
\footnote{Note that the individual migration times $T_{m,1}$ and $T_{m,2}$ enter the equations
only through the ``effective'' migration time $T_m\equiv (-T_{m,1}^{-1}+T_{m,2}^{-1})^{-1}$.}.
Since the first-order MMR Hamiltonian can be reduced to one degree of freedom 
(e.g., \citealt{Sessin84,Wisdom86}),
an analytical criterion (which is proved to be
qualitatively accurate by numerical results) can be obtained. \citet{DeckBatygin15}
found that the capture is stable when
\eq{
\left(1+\frac{q}{q_e}\right)^{-5/2}\left(1-\frac{q^2\alpha_{\rm res}}{R^2q_e}\right)\left(\frac{T_{e,1}}{T_m}\right)^{3/2}\lesssim \mu,
}
where $R$ is a parameter of order unity. Approximately, the above
relation is satisfied when $(T_{e,1}/T_m)^{1/2}\lesssim \mu^{1/3}$
[same as the result in \citet{GoldreichSchlichting14}] or $q\gtrsim
q_e^{1/2}$ (similar to our result for the $j\geq 5$ second-order MMR). 
Thus, the parameter space allowing stable first-order MMR
capture is larger than that allowing stable second-order MMR capture.

\bibliographystyle{mnras}
\bibliography{BIB}

\end{document}